\documentclass[10pt,twoside]{article}
\usepackage{amssymb}

\usepackage{graphicx}

\begin{document}

\title{BETOCS using the 157 gold SNe Ia Data : Hubble is not humble}
\author{R. Colistete Jr.\thanks{%
e-mail: \texttt{colistete@cce.ufes.br}} and R. Giostri\thanks{%
e-mail: \texttt{ramon\_giostri@yahoo.com.br}} \\
\\
\mbox{\small Universidade Federal do Esp\'{\i}rito Santo,
Departamento
de F\'{\i}sica}\\
\mbox{\small Av. Fernando Ferrari, 514, Campus de Goiabeiras, CEP
29075-910, Vit\'oria, Esp\'{\i}rito Santo, Brazil}}
\date{\today}
\maketitle

\begin{abstract}
The type Ia supernovae observational data is one of the most important in
observational cosmology nowadays. Here we present the first public version
of \textbf{BETOCS} (\textbf{B}ay\textbf{E}sian \textbf{T}ools for \textbf{O}%
bservational \textbf{C}osmology using \textbf{S}Ne Ia), which is a powerful
and high productivity tool aimed to help the theoretical physicist community
investigate cosmological models using type Ia supernovae (SNe Ia)
observational data. BETOCS is applied to the generalized Chaplygin gas model
(GCGM), traditional Chaplygin gas model (CGM) and $\Lambda$CDM, ranging from
5 to 3 free parameters, respectively. The ``gold sample'' of 157 supernovae
data is used. It is shown that the Chaplygin gas scenario is viable (in most
cases the $\Lambda$CDM is disfavoured) and the quartessence scenario (that
unifies the description for dark matter and dark energy) is favoured. The
Hubble parameter ($H_0$) is important and should not be fixed and it can be
estimated or marginalized with or without the Hubble Space Telescope prior. 
\vspace{0.7cm}
\end{abstract}

PACS number(s): 98.80.Bp, 98.80.Es, 04.60.Gw \vspace{0.7cm}

\section{Introduction}

The type Ia supernovae (SNe Ia) observational data has forced us to discard
or change the majority of theoretical cosmological models supposed to be
correct until the second half of the last decade \cite{riess,mutter}. The
crossing of the SNe Ia statistics with other observational data, like the
anisotropy of the cosmic microwave background radiation (CMBR) \cite{spergel}%
, gravitational lensing \cite{turner,fukugita}, the X-ray gas mass fraction
of galaxy clusters \cite{allen04}, etc, leads to a scenario where the matter
content of the Universe is described by an unclustered component of negative
pressure, the dark energy, and a clustered component of zero pressure, the
cold dark matter.

There are many candidate for dark energy, the most natural seems to be the
cosmological constant \cite{bagla}, since it can be connected with the
vacuum energy in quantum field theory \cite{weinberg}, but the small value
resulting from observations for the energy density of the cosmological
constant term yields a discrepancy of about $120$ orders of magnitude with
the theoretically predicted value \cite{carroll}. Among many other
possibilities, for example there is the quintessence model with scalar
fields \cite{stein1,stein2}.

Here we will focus on the Chaplygin gas models (CGM) \cite{kamenshchik,
maklerthesis, bilic2002, bento2002}. It is based on a string inspired
configuration that leads to a specific equation of state where pressure is
negative and varies with the inverse of the density \cite{jack}. This model
has been generalized, giving birth to the generalized Chaplygin gas model
(GCGM), where now the pressure varies with a power of the inverse of the
density \cite{bento2002}. These proposals have many advantages, among which
we can quote the following: in spite of presenting a negative pressure, the
sound velocity is positive, what assures stability \cite{fabris}; these
models can unify the description of dark energy and dark matter, since the
fluid can clusters at small scale, remaining a smooth component at large
scales \cite{bento2002}; the CGM has an interesting connection with string
theory \cite{jack}. Some criticisms have been addressed to the GCGM (CGM)
mainly connected with its features related to the power spectrum for the
agglomerated matter \cite{sand}. However, in our opinion, this specific
criticism is not conclusive, since the introduction of baryons may alleviate
the objections presented against the cosmological scenarios based on the
GCGM (GCM) \cite{avelino}.

The GCGM (Generalized Chaplygin Gas Model) is defined as a perfect fluid
with an equation of state given by 
\begin{equation}
p=-\frac{A}{\rho ^{\alpha }}\quad ,
\end{equation}
where $A$ and $\alpha $ are constants. When $\alpha =1$ we re-obtain the
equation of state for the CGM (Chaplygin Gas Model), the traditional
Chaplygin gas model. See refs. \cite{colistete3, colistete1, colistete2} for
more detailed definitions of the Chaplygin gas models used in the present
work.

All free parameters for each model are considered. In the case of the GCGM
there are five free parameter: the Hubble constant $H_{0}$; the equation of
state parameter $\alpha $; the ``sound velocity'' related parameter $\bar{A}$%
; the density parameter for the pressureless matter $\Omega _{m0}$; the
density parameter for the Chaplygin gas $\Omega _{c0}$ (or alternatively the
density parameter for the curvature density of the Universe $\Omega _{k0}$).
For the CGM, the number of parameters reduce to four, since $\alpha =1$. We
also consider the $\Lambda $CDM, where the number of parameters reduce to
three: $H_{0}$, $\Omega _{m0}$ and $\Omega _{c0}$ (alternatively, $\Omega
_{k0}$).

One important point is how to perform this statistical analysis: the final
conclusions may, in some cases, depend on the statistical framework
(Bayesian, frequentist, etc.), as well as on the parameters that are allowed
to be free, and how these parameters are constrained (through a joint
probability for two parameters, minimizing the error function or through a
marginalization of all parameters excepted one, etc.). In some cases, the
different procedures adopted may lead to quite different conclusions on the
best value for a given set of parameters. The choice of the observational
data sample may of course be important as well.

The present work is intended to :

\begin{enumerate}
\item  Announce the first public version of \textbf{BETOCS} (\textbf{B}ay%
\textbf{E}sian \textbf{T}ools for \textbf{O}bservational \textbf{C}osmology
using \textbf{S}Ne Ia \cite{betocs}), which is a powerful and high
productivity tool aimed to help the theoretical physicist community
investigate cosmological models using type Ia supernovae (SNe Ia)
observational data. BETOCS is a freeware and open source tool written in the 
\textit{Mathematica} \cite{mathematica} language. The \textit{Mathematica}
notebooks of BETOCS contain documentation, source code and practical
examples with textual and graphical outputs.

\item  Emphasize that fixing $H_{0}$ is not acceptable, yielding arbitrary
parameter estimations and usually bad best-fittings. On the other hand, the
HST (Hubble Space Telescope) prior \cite{freedman} for $H_{0}$ implies minor
effects on all best-fittings and parameter estimations (as shown here
comparing the tables with flat and HST priors for $H_{0}$), so its use is
just a matter of choice. Nevertheless, it is recommended to use the HST
prior as it is standard when using other observational cosmological data
(X-ray gas mass fraction, etc).

\item  Continue the work of refs. \cite{colistete3, colistete1, colistete2}
and show a very complete view and analysis of the Chaplygin gas models
(generalized and traditional) using SNe Ia. The present work specially fixes
an error on ref. \cite{colistete3} when calculating the best-fittings and
parameter estimations for non-flat Universes ( $\Omega _{k0}\neq 0$), due to
an earlier version of BETOCS with wrong optimization code. B$\frac{{}}{{}}$%
ut the side-effects of this error were not critical, just worsening some
positive features.
\end{enumerate}

The best-fitting parameters and the parameter estimations were calculated
for each case of GCGM, CGM or $\Lambda$CDM using a corresponding \textbf{%
BETOCS} (\textbf{B}ay\textbf{E}sian \textbf{T}ools for \textbf{O}%
bservational \textbf{C}osmology using \textbf{S}Ne Ia \cite{betocs})
notebook (written in \textit{Mathematica} \cite{mathematica} language)
containing the : definitions of the theoretical cosmological model, reading
of the observational SNe Ia data, $\chi ^2$ definition, Bayesian tools
library, calculation of the Bayesian PDF, global maximization of PDF, PDF
visualization in $3$ dimensions (if available), PDF visualization and
analysis in $2$ dimensions and finally PDF visualization in $1$ dimension
with parameter estimation. This article does not include all the graphics
and analyses of the BETOCS notebooks, but some of them are available on the
Internet site of the BETOCS project \cite{betocs}.

This paper is organized as follows. In section \ref{sectionBestfitting} we
detail the best-fitting analysis using BETOCS, with many results presented
in tables. Section \ref{sectionParameterEstimation} explains how the
parameter estimations are made using BETOCS, such that a detailed Bayesian
analysis is performed for each independent and dependent parameter, with the
results shown in many tables and figures. The conclusions are discussed in
section \ref{sectionConclusions}.

\section{Best-fitting parameters using BETOCS}

\label{sectionBestfitting}

In order to compare the theoretical results with the observational data, the
first step in this sense is to compute the quality of the fitting through
the least squared fitting quantity $\chi ^{2}$. In the case of flat priors
for all independent parameters of the theoretical cosmological model, we get 
\begin{equation}
\chi ^{2}=\sum_{i}\frac{\left( \mu _{0,i}^{o}-\mu _{0,i}^{t}\right) ^{2}}{%
\sigma _{\mu _{0},i}^{2}}\quad .  \label{Chi2}
\end{equation}
In this expression, $\mu _{0,i}^{o}$ is the distance moduli observationaly
measured for each supernova of the $157$ gold SNe Ia dataset \cite{riessa}, $%
\mu _{0,i}^{t}$ is the value calculated through the theoretical cosmological
model, $\sigma _{\mu _{0},i}^{2}$ is the measurement error and includes the
dispersion in the distance modulus due to the dispersion in galaxy redshift
due to peculiar velocities, following ref. \cite{riessa}. It is useful to
define $\chi _{\nu }^{2}$ : $\chi ^{2}$ divided by the number of degrees of
freedom of the observational data, i.e., the number of SNe Ia, here $157$.

As we also want to compare the fitting and estimation of the parameters
without priors and with the HST (Hubble Space Telescope) prior \cite
{freedman} for $H_{0}$, then the $\chi ^{2}$ used for the calculations with
the HST prior is simply 
\begin{equation}
\chi ^{2}=\sum_{i}\frac{\left( \mu _{0,i}^{o}-\mu _{0,i}^{t}\right) ^{2}}{%
\sigma _{\mu _{0},i}^{2}}+\frac{(H_{0}-72)^{2}}{8^{2}}\quad .
\label{Chi2HST}
\end{equation}

\begin{table}[ht!]
\begin{center}
\begin{tabular}{|c|c|c|c|c|c|c|}
\hline\hline
&  &  &  &  &  &  \\[-7pt] 
GCGM &  &  &  &  & $k=0$, & $k=0$, \\ 
with &  & $k=0$ & $\Omega _{m0}=0$ & $\Omega _{m0}=0.04$ & $\Omega _{m0}=0$
& $\Omega _{m0}=0.04$ \\[2pt] \hline
&  &  &  &  &  &  \\[-7pt] 
$\chi _{\nu }^{2}$ & $1.1075$ & $1.1094$ & $1.1075$ & $1.1081$ & $1.1094$ & $%
1.1096$ \\[2pt] \hline
&  &  &  &  &  &  \\[-7pt] 
$\alpha $ & $7.70$ & $2.83$ & $7.53$ & $7.75$ & $2.83$ & $3.07$ \\%
[2pt] \hline
&  &  &  &  &  &  \\[-7pt] 
$H_{0}$ & $65.03$ & $65.12$ & $65.03$ & $64.98$ & $65.12$ & $65.09$ \\%
[2pt] \hline
&  &  &  &  &  &  \\[-7pt] 
$\Omega _{k0}$ & $0.177$ & $0$ & $0.173$ & $0.153$ & $0$ & $0$ \\[2pt] \hline
&  &  &  &  &  &  \\[-7pt] 
$\Omega _{m0}$ & $0.000$ & $0.000$ & $0$ & $0.04$ & $0$ & $0.04$ \\%
[2pt] \hline
&  &  &  &  &  &  \\[-7pt] 
$\Omega _{c0}$ & $0.823$ & $1.000$ & $0.827$ & $0.807$ & $1$ & $0.96$ \\%
[2pt] \hline
&  &  &  &  &  &  \\[-7pt] 
$\bar{A}$ & $0.996$ & $0.929$ & $0.996$ & $0.998$ & $0.929$ & $0.949$ \\%
[2pt] \hline
&  &  &  &  &  &  \\[-7pt] 
$t_{0}$ & $13.53$ & $13.55$ & $13.53$ & $13.53$ & $13.55$ & $13.55$ \\%
[2pt] \hline
&  &  &  &  &  &  \\[-7pt] 
$q_{0}$ & $-0.819$ & $-0.893$ & $-0.822$ & $-0.785$ & $-0.893$ & $-0.866$ \\%
[2pt] \hline
&  &  &  &  &  &  \\[-7pt] 
$a_{i}$ & $0.786$ & $0.753$ & $0.786$ & $0.777$ & $0.753$ & $0.751$ \\%
[2pt] \hline\hline
\end{tabular}
\end{center}
\caption{The best-fitting parameters, i.e., when $\protect\chi _{\protect\nu
}^{2}$ is minimum, for each type of spatial section and matter content of
the generalized Chaplygin gas model. $H_{0}$ is given in $km/M\!pc.s$, $\bar{%
A}$ in units of $c$, $t_{0}$ in $Gy$ and $a_{i}$ in units of $a_{0}$. }
\label{tableBestFitGCG}
\end{table}

\begin{table}[ht!]
\begin{center}
\begin{tabular}{|c|c|c|c|c|c|c|}
\hline\hline
&  &  &  &  &  &  \\[-7pt] 
GCGM &  &  &  &  & $k=0$, & $k=0$, \\ 
with &  & $k=0$ & $\Omega _{m0}=0$ & $\Omega _{m0}=0.04$ & $\Omega _{m0}=0$
& $\Omega _{m0}=0.04$ \\[2pt] \hline
&  &  &  &  &  &  \\[-7pt] 
$\chi _{\nu }^{2}$ & $1.1122$ & $1.1140$ & $1.1122$ & $1.1129$ & $1.1140$ & $%
1.1143$ \\[2pt] \hline
&  &  &  &  &  &  \\[-7pt] 
$\alpha $ & $7.73$ & $2.96$ & $7.27$ & $7.57$ & $2.96$ & $3.22$ \\%
[2pt] \hline
&  &  &  &  &  &  \\[-7pt] 
$H_{0}$ & $65.11$ & $65.20$ & $65.11$ & $65.06$ & $65.20$ & $65.17$ \\%
[2pt] \hline
&  &  &  &  &  &  \\[-7pt] 
$\Omega _{k0}$ & $0.170$ & $0$ & $0.162$ & $0.143$ & $0$ & $0$ \\[2pt] \hline
&  &  &  &  &  &  \\[-7pt] 
$\Omega _{m0}$ & $0.000$ & $0.000$ & $0$ & $0.04$ & $0$ & $0.04$ \\%
[2pt] \hline
&  &  &  &  &  &  \\[-7pt] 
$\Omega _{c0}$ & $0.830$ & $1.000$ & $0.838$ & $0.817$ & $1$ & $0.96$ \\%
[2pt] \hline
&  &  &  &  &  &  \\[-7pt] 
$\bar{A}$ & $0.996$ & $0.935$ & $0.995$ & $0.997$ & $0.935$ & $0.954$ \\%
[2pt] \hline
&  &  &  &  &  &  \\[-7pt] 
$t_{0}$ & $13.53$ & $13.54$ & $13.52$ & $13.53$ & $13.54$ & $13.54$ \\%
[2pt] \hline
&  &  &  &  &  &  \\[-7pt] 
$q_{0}$ & $-0.826$ & $-0.902$ & $-0.832$ & $-0.794$ & $-0.902$ & $-0.874$ \\%
[2pt] \hline
&  &  &  &  &  &  \\[-7pt] 
$a_{i}$ & $0.785$ & $0.754$ & $0.784$ & $0.776$ & $0.754$ & $0.752$ \\%
[2pt] \hline\hline
\end{tabular}
\end{center}
\caption{The best-fitting parameters, i.e., when $\protect\chi _{\protect\nu
}^{2}$ is minimum, for each type of spatial section and matter content of
the generalized Chaplygin gas model using the HST prior. $H_{0}$ is given in 
$km/M\!pc.s$, $\bar{A}$ in units of $c$, $t_{0}$ in $Gy$ and $a_{i}$ in
units of $a_{0}$. }
\label{tableBestFitGCGp}
\end{table}

\begin{table}[ht!]
\begin{center}
\begin{tabular}{|c|c|c|c|c|c|c|}
\hline\hline
&  &  &  &  &  &  \\[-7pt] 
CGM &  &  &  &  & $k=0$, & $k=0$, \\ 
with &  & $k=0$ & $\Omega _{m0}=0$ & $\Omega _{m0}=0.04$ & $\Omega _{m0}=0$
& $\Omega _{m0}=0.04$ \\[2pt] \hline
&  &  &  &  &  &  \\[-7pt] 
$\chi _{\nu }^{2}$ & $1.1119$ & $1.1141$ & $1.1119$ & $1.1121$ & $1.1141$ & $%
1.1147$ \\[2pt] \hline
&  &  &  &  &  &  \\[-7pt] 
$H_{0}$ & $64.96$ & $64.73$ & $64.96$ & $64.95$ & $64.73$ & $64.70$ \\%
[2pt] \hline
&  &  &  &  &  &  \\[-7pt] 
$\Omega _{k0}$ & $-0.149$ & $0$ & $-0.149$ & $-0.160$ & $0$ & $0$ \\%
[2pt] \hline
&  &  &  &  &  &  \\[-7pt] 
$\Omega _{m0}$ & $0.000$ & $0.000$ & $0$ & $0.04$ & $0$ & $0.04$ \\%
[2pt] \hline
&  &  &  &  &  &  \\[-7pt] 
$\Omega _{c0}$ & $1.149$ & $1.000$ & $1.149$ & $1.120$ & $1$ & $0.96$ \\%
[2pt] \hline
&  &  &  &  &  &  \\[-7pt] 
$\bar{A}$ & $0.806$ & $0.811$ & $0.806$ & $0.825$ & $0.811$ & $0.834$ \\%
[2pt] \hline
&  &  &  &  &  &  \\[-7pt] 
$t_{0}$ & $13.91$ & $14.01$ & $13.91$ & $13.92$ & $14.01$ & $14.03$ \\%
[2pt] \hline
&  &  &  &  &  &  \\[-7pt] 
$q_{0}$ & $-0.815$ & $-0.717$ & $-0.815$ & $-0.807$ & $-0.717$ & $-0.701$ \\%
[2pt] \hline
&  &  &  &  &  &  \\[-7pt] 
$a_{i}$ & $0.702$ & $0.699$ & $0.702$ & $0.699$ & $0.699$ & $0.695$ \\%
[2pt] \hline\hline
\end{tabular}
\end{center}
\caption{The best-fitting parameters, i.e., when $\protect\chi _{\protect\nu
}^{2}$ is minimum, for each type of spatial section and matter content of
the traditional Chaplygin gas model. $H_{0}$ is given in $km/M\!pc.s$, $\bar{%
A}$ in units of $c$, $t_{0}$ in $Gy$ and $a_{i}$ in units of $a_{0}$. }
\label{tableBestFitCG}
\end{table}

\begin{table}[ht!]
\begin{center}
\begin{tabular}{|c|c|c|c|c|c|c|}
\hline\hline
&  &  &  &  &  &  \\[-7pt] 
CGM &  &  &  &  & $k=0$, & $k=0$, \\ 
with &  & $k=0$ & $\Omega _{m0}=0$ & $\Omega _{m0}=0.04$ & $\Omega _{m0}=0$
& $\Omega _{m0}=0.04$ \\[2pt] \hline
&  &  &  &  &  &  \\[-7pt] 
$\chi _{\nu }^{2}$ & $1.1167$ & $1.1193$ & $1.1167$ & $1.1170$ & $1.1193$ & $%
1.1200$ \\[2pt] \hline
&  &  &  &  &  &  \\[-7pt] 
$H_{0}$ & $65.04$ & $64.80$ & $65.04$ & $65.02$ & $64.80$ & $64.76$ \\%
[2pt] \hline
&  &  &  &  &  &  \\[-7pt] 
$\Omega _{k0}$ & $-0.158$ & $0$ & $-0.158$ & $-0.170$ & $0$ & $0$ \\%
[2pt] \hline
&  &  &  &  &  &  \\[-7pt] 
$\Omega _{m0}$ & $0.000$ & $0.000$ & $0$ & $0.04$ & $0$ & $0.04$ \\%
[2pt] \hline
&  &  &  &  &  &  \\[-7pt] 
$\Omega _{c0}$ & $1.158$ & $1.000$ & $1.158$ & $1.130$ & $1$ & $0.96$ \\%
[2pt] \hline
&  &  &  &  &  &  \\[-7pt] 
$\bar{A}$ & $0.808$ & $0.813$ & $0.808$ & $0.827$ & $0.813$ & $0.836$ \\%
[2pt] \hline
&  &  &  &  &  &  \\[-7pt] 
$t_{0}$ & $13.91$ & $14.02$ & $13.91$ & $13.93$ & $14.02$ & $14.04$ \\%
[2pt] \hline
&  &  &  &  &  &  \\[-7pt] 
$q_{0}$ & $-0.824$ & $-0.720$ & $-0.824$ & $-0.816$ & $-0.720$ & $-0.703$ \\%
[2pt] \hline
&  &  &  &  &  &  \\[-7pt] 
$a_{i}$ & $0.701$ & $0.697$ & $0.701$ & $0.698$ & $0.697$ & $0.693$ \\%
[2pt] \hline\hline
\end{tabular}
\end{center}
\caption{The best-fitting parameters, i.e., when $\protect\chi _{\protect\nu
}^{2}$ is minimum, for each type of spatial section and matter content of
the traditional Chaplygin gas model using the HST prior. $H_{0}$ is given in 
$km/M\!pc.s$, $\bar{A}$ in units of $c$, $t_{0}$ in $Gy$ and $a_{i}$ in
units of $a_{0}$. }
\label{tableBestFitCGp}
\end{table}

\begin{table}[th]
\begin{center}
\begin{tabular}{|c|c|c|c|c|c|c|c|c|}
\hline\hline
& \multicolumn{2}{|c|}{} & \multicolumn{2}{|c|}{} & \multicolumn{2}{|c|}{} & 
\multicolumn{2}{|c|}{} \\[-7pt] 
& \multicolumn{2}{|c|}{$\Lambda CDM$} & \multicolumn{2}{|c|}{$\Lambda CDM$ :}
& \multicolumn{2}{|c|}{$\Lambda CDM$ :} & \multicolumn{2}{|c|}{$\Lambda CDM$
:} \\ 
& \multicolumn{2}{|c|}{} & \multicolumn{2}{|c|}{$k=0$} & 
\multicolumn{2}{|c|}{$\Omega _{m0}=0$} & \multicolumn{2}{|c|}{$\Omega
_{m0}=0.04$} \\[2pt] \hline
&  &  &  &  &  &  &  &  \\[-7pt] 
$H_{0}$ Prior & $flat$ & $HST$ & $flat$ & $HST$ & $flat$ & $HST$ & $flat$ & $%
HST$ \\[2pt] \hline
&  &  &  &  &  &  &  &  \\[-7pt] 
$\chi _{\nu }^{2}$ & $1.1149$ & $1.1199$ & $1.1279$ & $1.1337$ & $1.2002$ & $%
1.2068$ & $1.1882$ & $1.1946$ \\[2pt] \hline
&  &  &  &  &  &  &  &  \\[-7pt] 
$H_{0}$ & $64.85$ & $64.93$ & $64.32$ & $64.39$ & $63.78$ & $63.86$ & $63.90$
& $63.98$ \\[2pt] \hline
&  &  &  &  &  &  &  &  \\[-7pt] 
$\Omega _{k0}$ & $-0.437$ & $-0.447$ & $0$ & $0$ & $0.803$ & $0.795$ & $%
0.692 $ & $0.684$ \\[2pt] \hline
&  &  &  &  &  &  &  &  \\[-7pt] 
$\Omega _{m0}$ & $0.459$ & $0.460$ & $0.309$ & $0.306$ & $0$ & $0$ & $0.04$
& $0.04$ \\[2pt] \hline
&  &  &  &  &  &  &  &  \\[-7pt] 
$\Omega _{\Lambda}$ & $0.978$ & $0.987$ & $0.691$ & $0.694$ & $0.197$ & $%
0.205$ & $0.268$ & $0.276$ \\[2pt] \hline
&  &  &  &  &  &  &  &  \\[-7pt] 
$t_{0}$ & $14.83$ & $14.85$ & $14.87$ & $14.88$ & $16.85$ & $16.88$ & $16.10$
& $16.13$ \\[2pt] \hline
&  &  &  &  &  &  &  &  \\[-7pt] 
$q_{0}$ & $-0.749$ & $-0.757$ & $-0.537$ & $-0.540$ & $-0.197$ & $-0.205$ & $%
-0.248$ & $-0.256$ \\[2pt] \hline
&  &  &  &  &  &  &  &  \\[-7pt] 
$a_{i}$ & $0.617$ & $0.615$ & $0.607$ & $0.605$ & $0$ & $0$ & $0.421$ & $%
0.417$ \\[2pt] \hline\hline
\end{tabular}
\end{center}
\caption{The best-fitting parameters, i.e., when $\protect\chi _{\protect\nu
}^{2}$ is minimum, for each type of spatial section and matter content of
the $\Lambda $CDM model using flat and HST priors. $H_{0}$ is given in $%
km/M\!pc.s$, $t_{0}$ in $Gy$ and $a_{i}$ in units of $a_{0}$. }
\label{tableBestFitLCDM}
\end{table}

In tables \ref{tableBestFitGCG} and \ref{tableBestFitGCGp} the values of the
parameters for the minimum $\chi _{\nu }^{2}$ ($\chi^{2}$ divided by the
number of SNe Ia) are given for the GCGM with five free parameters $(\alpha
,H_{0},\Omega _{m0},\Omega _{c0},\bar{A})$ and for other cases where the
pressureless mater, the curvature or both are fixed, respectively using the
flat prior for $H_{0}$ and the HST (Hubble Space Telescope) prior for $H_{0}$%
. Analogously, the same estimations are presented in tables \ref
{tableBestFitCG} and \ref{tableBestFitCGp} for the CGM, for up to four free
parameters $(H_{0},\Omega _{m0},\Omega _{c0},\bar{A})$, and in table \ref
{tableBestFitLCDM} for the $\Lambda $CDM, for up to three free parameters $%
(H_{0},\Omega _{m0},\Omega _{\Lambda })$.

It is important to emphasize that, for each case, all free independent
parameters are considered simultaneously to obtain the minimum of $\chi
_{\nu }^{2}$. So, for example assuming the GCGM, if we ask for the best
simultaneous values of $(\alpha ,H_{0},\Omega _{m0},\Omega _{c0},\bar{A})$
then the answer is given by the first column of table \ref{tableBestFitGCG}.
However, in this example, asking for the best value of $\alpha $ by weighing
(marginalizing or integrating) all possible values of $(H_{0},\Omega
_{m0},\Omega _{c0},\bar{A})$ yields the estimation in the first column of
table \ref{tableParEstGCG},\ whose peak of $0.59$ for $\alpha $ is totally
different from $7.70$ as best-fitting parameter ! The parameter estimation
issue is addressed by the Bayesian statistics of the next section, not by
best-fitting in $n$-dimensional parameter space.

Each column of these best-fitting tables was calculated using a
corresponding BETOCS \cite{betocs} notebook. The minimization of $\chi _{\nu
}^{2}$ is obtained in the following way : the initial global minimum taken
from the $n$-dimensional discrete parameter space (see next section) is used
as initial value to search for the local minimum of $\chi _{\nu }^{2}$ by
using the function \textit{FindMinimum} of the software \textit{Mathematica} 
\cite{mathematica}.

Note that the minimum values for $\chi ^{2}$ using the ``gold sample'' are
worse (i.e., higher, from $1.11$ to $1.20$) than the corresponding ones
(between $0.74$ and $0.77$) using the restricted sample of $26$ supernovae 
\cite{colistete1,colistete2} (which have excellent quality). While the
minima for $\chi ^{2}$ are always slightly higher when the HST prior is used
(because the HST prior peak is far from the best-fitting $H_{0}$), yielding
best-fitting parameter values with minor changes. Other important results :
the parameter $\alpha $ is usually much bigger than $1$, $\bar{A}$ is often
near unity, $\Omega _{k0}$ is not far from $0$ suggesting a flat spatial
section (except for $\Lambda $CDM), $\Omega _{m0}$ being null recovers the
quartessence \cite{maklerthesis, makler2003} scenario (except for $\Lambda $%
CDM), $q_{0}$ and $a_{i}$ point to an accelerating Universe today with an
age $t_{0}$ of approximately $14$ Gy. But these results must be compared
with a more complete statistical analysis to be presented below.

\section{Parameter estimations using BETOCS}

\label{sectionParameterEstimation}

Following the previous works \cite{colistete3, colistete1, colistete2}, the
Bayesian statistical analysis is employed here instead of the more usual
frequentist (or standard or traditional) statistics. The Bayesian statistics
emphasizes considering only the (observational) data you have, rather than
simulating an infinite space of data, which is an advantage. On the other
hand, the Bayesian marginalization process is computationally time-consuming
if the number of parameters of the theoretical model is large. For the case
here, with a maximum of five free parameters and low number of data points
(157 SNe Ia), the Bayesian approach is better suited than the frequentist
statistics. See Refs. \cite{Loredo,Loredo2,Gregory,Abroe} for discussions
about the frequentist versus Bayesian statistics and some applications in
physics.

The probability of the set of distance moduli $\mu _{0}$ conditional on the
values of a set of parameters $\{p_{i}\}$ is given by the Gaussian : 
\begin{equation}
p(\mu _{0}|\{p_{i}\})\propto \exp \biggr(-\frac{\chi ^{2}}{2\,}\biggl)\quad .
\label{pd1}
\end{equation}
This probability distribution must be normalized. Evidently, when, for a set
of values of the parameters, the $\chi ^{2}$ is minimum the probability is
maximum. This is a valuable information but is not enough to constraint the
parameters.

From the probability distribution (\ref{pd1}), a joint probability
distribution for any subset of parameters can be obtained by integrating
(marginalizing) on the remaining parameters, see refs. \cite{colistete1,
colistete2}. So, in order to properly estimate a single parameter, the
probability distribution must be marginalized on all other parameters,
usually yielding a quite different result if we try to estimate the
parameter in a two or three-dimensional parameter space. The reason is that,
in such multidimensional parameter space, if a parameter has a large
probability density but in a narrow region, the total contribution of this
region may be quite small compared to other large regions which have small
probability: in the marginalization process, this kind of high PDF region
contributes little to the estimation of a given parameter.

Hence the estimation of a given parameter will be made by marginalizing on
all other ones. A detailed Bayesian analysis of the independent and
dependent parameters is shown in tables \ref{tableParEstGCG}, \ref
{tableParEstCG} and \ref{tableParEstLCDM} for GCGM, CGM and $\Lambda $CDM
with flat prior for $H_{0}$, and in tables \ref{tableParEstGCGp}, \ref
{tableParEstCGp} and \ref{tableParEstLCDMp} for GCGM, CGM and $\Lambda $CDM
with the HST (Hubble Space Telescope) prior for $H_{0}$. Each column of
these tables was calculated using a corresponding BETOCS \cite{betocs}
notebook (including the best-fitting calculations for the specific model).

The following estimation analyses will focus on the tables \ref
{tableParEstGCG}--\ref{tableParEstLCDMp} and the accompanying figures.

\subsection{Estimation of $H_0$ : Hubble is not humble}

The predicted value of the Hubble constant today $H_{0}$ is the most robust
one, with minor changes for the different models (GCGM, CGM and $\Lambda $%
DCM) and cases of fixed parameters. When comparing with ref. \cite
{colistete3}, $H_{0}$ is now slightly smaller.

\begin{figure}[th]
\begin{center}
\includegraphics[trim=0.4in 0.2in 0.2in 0in,scale=0.7]
{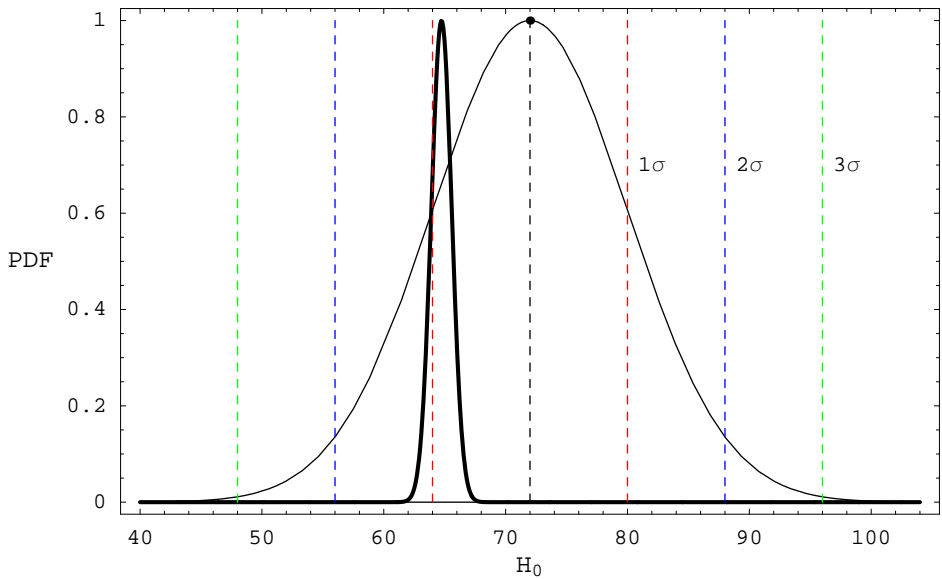}
\end{center}
\caption{{\protect\footnotesize The one-dimensional PDF for $H_{0}$. The
thin line shows the HST (Hubble Space Telescope) prior, with the $1\protect%
\sigma $ ($68.27\%$), $2\protect\sigma $ ($95.45\%$) and $3\protect\sigma $ (%
$99.73\%$) regions delimited by red, blue and green lines, respectively. The
thick line shows a typical $H_{0}$ estimation from SNe Ia analysis in this
article, clearly showing that the HST prior has much larger dispersion and
its $2\protect\sigma $ region includes the $H_{0}$ estimation from SNe Ia.}}
\label{figH0prior}
\end{figure}

Figure \ref{figH0prior} shows a typical $H_{0}$ estimation from SNe Ia
analysis in this article, and the HST (Hubble Space Telescope) prior \cite
{freedman} for $H_{0}$, which has a much larger dispersion. The effect of
the HST prior on the $H_{0}$ estimation is small : it slightly changes the
shape PDF for $H_{0}$, the PDF peak moves increases (moves to the right) and
there are some very \ small changes on the left and right dispersions.

The minor effect of the HST prior can also be verified by comparing tables 
\ref{tableParEstGCG}, \ref{tableParEstCG} and \ref{tableParEstLCDM} for
GCGM, CGM and $\Lambda$CDM with flat prior for $H_0$ with tables \ref
{tableParEstGCGp}, \ref{tableParEstCGp} and \ref{tableParEstLCDMp} for GCGM,
CGM and $\Lambda$CDM with the HST (Hubble Space Telescope) prior for $H_0$.

It is important to emphasize that fixing $H_{0}$ is not acceptable as the
n-dimensional PDF quite depends on the $H_{0}$ parameter. For $\Lambda $CDM
as an example : $\chi _{\nu }^{2}$ is very high ($1.563$), $\Omega
_{k0}=-0.890_{-0.241}^{+0.286}$ and many other totally different parameter
estimations.

\subsection{Estimation of $\protect\alpha$}

With respect to ref. \cite{colistete3}, the peak values of $\alpha$ are
slightly increased and the dispersion is also a little larger, for example,
the GCGM now gives $\alpha =-0.59_{-0.41}^{+5.27}$. Imposing that the space
is flat or fixing the pressureless matter lead to positive best values for $%
\alpha$. For example, the quartessence \cite{maklerthesis, makler2003}
scenario ($\Omega_m = 0$) predicts $\alpha = 0.90^{+5.52}_{-1.83}$.

\begin{table}[t!]
\begin{center}
\begin{tabular}{|c|c|c|c|c|c|c|}
\hline\hline
&  &  &  &  &  &  \\[-7pt] 
GCGM &  &  &  &  & $k=0$, & $k=0$, \\ 
with &  & $k=0$ & $\Omega _{m0}=0$ & $\Omega _{m0}=0.04$ & $\Omega _{m0}=0$
& $\Omega _{m0}=0.04$ \\[2pt] \hline
&  &  &  &  &  &  \\[-7pt] 
$\alpha$ & $-0.59_{-0.41}^{+5.27}$ & $1.18_{-2.18}^{+4.22}$ & $%
0.90_{-1.83}^{+5.52}$ & $0.64_{-1.64}^{+5.55}$ & $1.57_{-1.95}^{+5.09}$ & $%
1.52_{-1.97}^{+4.76}$ \\[2pt] \hline
&  &  &  &  &  &  \\[-7pt] 
$H_{0}$ & $64.68_{-1.69}^{+1.73}$ & $64.52_{-1.57}^{+1.64}$ & $%
64.73_{-1.72}^{+1.74}$ & $64.70_{-1.72}^{+1.75}$ & $64.93_{-1.66}^{+1.55}$ & 
$64.84_{-1.62}^{+1.54}$ \\[2pt] \hline
&  &  &  &  &  &  \\[-7pt] 
$\Omega_{k0}$ & $-0.251_{-0.694}^{+0.605}$ & $0$ & $0.065_{-0.833}^{+0.509}$
& $0.025_{-0.827}^{+0.496}$ & $0$ & $0$ \\[2pt] \hline
&  &  &  &  &  &  \\[-7pt] 
$\Omega_{m0}$ & $0.000_{-0.000}^{+0.499}$ & $0.000_{-0.000}^{+0.301}$ & $0$
& $0.04 $ & $0$ & $0.04$ \\[2pt] \hline
&  &  &  &  &  &  \\[-7pt] 
$\Omega_{c0}$ & $1.012_{-0.489}^{+0.667}$ & $1.000_{-0.301}^{+0.000}$ & $%
0.935_{-0.509}^{+0.833}$ & $0.935_{-0.496}^{+0.827}$ & $1$ & $0.96$ \\%
[2pt] \hline
&  &  &  &  &  &  \\[-7pt] 
$\bar{A}$ & $1.000_{-0.348}^{+0.000}$ & $0.989_{-0.245}^{+0.011}$ & $%
0.987_{-0.388}^{+0.012}$ & $0.988_{-0.382}^{+0.012}$ & $%
0.987_{-0.293}^{+0.013}$ & $0.987_{-0.278}^{+0.012}$ \\[2pt] \hline
&  &  &  &  &  &  \\[-7pt] 
$t_{0}$ & $14.42_{-1.77}^{+2.51}$ & $13.74_{-0.96}^{+1.76}$ & $%
13.77_{-1.08}^{+2.39}$ & $13.77_{-1.06}^{+2.82}$ & $13.56_{-0.83}^{+1.34}$ & 
$13.60_{-0.76}^{+1.40}$ \\[2pt] \hline
&  &  &  &  &  &  \\[-7pt] 
$q_{0}$ & $-0.730_{-0.328}^{+0.352}$ & $-0.717_{-0.247}^{+0.305}$ & $%
-0.750_{-0.399}^{+0.392}$ & $-0.750_{-0.372}^{+0.399}$ & $%
-0.928_{-0.114}^{+0.394}$ & $-0.886_{-0.106}^{+0.369}$ \\[2pt] \hline
&  &  &  &  &  &  \\[-7pt] 
$a_{i}$ & $0.626_{-0.123}^{+0.184}$ & $0.740_{-0.179}^{+0.073}$ & $%
0.735_{-0.262}^{+0.121}$ & $0.740_{-0.294}^{+0.108}$ & $%
0.760_{-0.147}^{+0.069}$ & $0.752_{-0.151}^{+0.072}$ \\[2pt] \hline
&  &  &  &  &  &  \\[-7pt] 
$p(\alpha>0)$ & $66.74\,\%$ & $87.82\,\%$ & $87.92\,\%$ & $85.40\,\%$ & $%
96.46\,\%$ & $95.61\,\%$ \\[2pt] \hline
&  &  &  &  &  &  \\[-7pt] 
$p(\alpha=1)$ & $40.81\,\%$ & $91.09\,\%$ & $95.94\,\%$ & $85.17\,\%$ & $%
70.44\,\%$ & $73.19\,\%$ \\[2pt] \hline
&  &  &  &  &  &  \\[-7pt] 
$p(\Omega_{k0}<0)$ & $79.05\,\%$ & $-$ & $49.21\,\%$ & $55.18\,\%$ & $-$ & $%
- $ \\[2pt] \hline
&  &  &  &  &  &  \\[-7pt] 
$p(\Omega_{k0}=0)$ & $45.70\,\%$ & $-$ & $82.48\,\%$ & $93.06\,\%$ & $-$ & $%
- $ \\[2pt] \hline
&  &  &  &  &  &  \\[-7pt] 
$p(\bar{A} \neq 1)$ & $0\,\%$ & $59.15\,\%$ & $100\,\%$ & $100\,\%$ & $%
100\,\%$ & $100\,\%$ \\[2pt] \hline
&  &  &  &  &  &  \\[-7pt] 
$p(q_{0}<0)$ & $5.14\,\sigma$ & $7.07\,\sigma$ & $100\,\%$ & $100\,\%$ & $%
100\,\%$ & $100\,\%$ \\[2pt] \hline
&  &  &  &  &  &  \\[-7pt] 
$p(a_{i}<1)$ & $5.26\,\sigma$ & $7.49\,\sigma$ & $100\,\%$ & $5.56\,\sigma$
& $100\,\%$ & $100\,\%$ \\[2pt] \hline\hline
\end{tabular}
\end{center}
\caption{The estimated parameters for the generalized Chaplygin gas model
(GCGM) and some specific cases of spatial section and matter content. We use
the Bayesian analysis to obtain the peak of the one-dimensional marginal
probability and the $2\,\protect\sigma $ credible region for each parameter. 
$H_{0}$ is given in $km/M\!pc.s$, $\bar{A}$ in units of $c$, $t_{0}$ in $Gy$
and $a_{i}$ in units of $a_{0}$.}
\label{tableParEstGCG}
\end{table}

\begin{table}[t!]
\begin{center}
\begin{tabular}{|c|c|c|c|c|c|c|}
\hline\hline
&  &  &  &  &  &  \\[-7pt] 
GCGM &  &  &  &  & $k=0$, & $k=0$, \\ 
with &  & $k=0$ & $\Omega _{m0}=0$ & $\Omega _{m0}=0.04$ & $\Omega _{m0}=0$
& $\Omega _{m0}=0.04$ \\[2pt] \hline
&  &  &  &  &  &  \\[-7pt] 
$\alpha$ & $-0.57_{-0.43}^{+5.23}$ & $1.24_{-2.24}^{+4.19}$ & $%
0.93_{-1.85}^{+5.46}$ & $0.67_{-1.66}^{+5.45}$ & $1.63_{-1.96}^{+5.04}$ & $%
1.57_{-1.98}^{+4.71}$ \\[2pt] \hline
&  &  &  &  &  &  \\[-7pt] 
$H_{0}$ & $64.78_{-1.70}^{+1.70}$ & $64.61_{-1.57}^{+1.63}$ & $%
64.83_{-1.74}^{+1.72}$ & $64.80_{-1.74}^{+1.72}$ & $65.00_{-1.65}^{+1.54}$ & 
$64.92_{-1.62}^{+1.52}$ \\[2pt] \hline
&  &  &  &  &  &  \\[-7pt] 
$\Omega_{k0}$ & $-0.261_{-0.687}^{+0.602}$ & $0$ & $0.055_{-0.825}^{+0.503}$
& $0.015_{-0.819}^{+0.491}$ & $0$ & $0$ \\[2pt] \hline
&  &  &  &  &  &  \\[-7pt] 
$\Omega_{m0}$ & $0.000_{-0.000}^{+0.499}$ & $0.000_{-0.000}^{+0.297}$ & $0$
& $0.04 $ & $0$ & $0.04$ \\[2pt] \hline
&  &  &  &  &  &  \\[-7pt] 
$\Omega_{c0}$ & $1.022_{-0.485}^{+0.661}$ & $1.000_{-0.297}^{+0.000}$ & $%
0.945_{-0.503}^{+0.825}$ & $0.945_{-0.491}^{+0.819}$ & $1$ & $0.96$ \\%
[2pt] \hline
&  &  &  &  &  &  \\[-7pt] 
$\bar{A}$ & $1.000_{-0.345}^{+0.000}$ & $0.989_{-0.240}^{+0.011}$ & $%
0.987_{-0.384}^{+0.012}$ & $0.988_{-0.378}^{+0.012}$ & $%
0.987_{-0.286}^{+0.013}$ & $0.987_{-0.271}^{+0.012}$ \\[2pt] \hline
&  &  &  &  &  &  \\[-7pt] 
$t_{0}$ & $14.43_{-1.78}^{+2.50}$ & $13.74_{-0.95}^{+1.76}$ & $%
13.77_{-1.09}^{+2.37}$ & $13.77_{-1.06}^{+2.80}$ & $13.48_{-0.66}^{+1.37}$ & 
$13.49_{-0.62}^{+1.45}$ \\[2pt] \hline
&  &  &  &  &  &  \\[-7pt] 
$q_{0}$ & $-0.741_{-0.324}^{+0.351}$ & $-0.717_{-0.251}^{+0.296}$ & $%
-0.757_{-0.400}^{+0.386}$ & $-0.764_{-0.366}^{+0.397}$ & $%
-0.931_{-0.112}^{+0.385}$ & $-0.888_{-0.104}^{+0.361}$ \\[2pt] \hline
&  &  &  &  &  &  \\[-7pt] 
$a_{i}$ & $0.623_{-0.120}^{+0.185}$ & $0.740_{-0.179}^{+0.073}$ & $%
0.733_{-0.257}^{+0.123}$ & $0.740_{-0.292}^{+0.105}$ & $%
0.760_{-0.144}^{+0.068}$ & $0.752_{-0.149}^{+0.071}$ \\[2pt] \hline
&  &  &  &  &  &  \\[-7pt] 
$p(\alpha>0)$ & $66.99\,\%$ & $88.41\,\%$ & $88.15\,\%$ & $85.64\,\%$ & $%
96.74\,\%$ & $95.92\,\%$ \\[2pt] \hline
&  &  &  &  &  &  \\[-7pt] 
$p(\alpha=1)$ & $40.97\,\%$ & $88.41\,\%$ & $96.96\,\%$ & $86.08\,\%$ & $%
67.80\,\%$ & $70.64\,\%$ \\[2pt] \hline
&  &  &  &  &  &  \\[-7pt] 
$p(\Omega_{k0}<0)$ & $80.04\,\%$ & $-$ & $50.58\,\%$ & $56.64\,\%$ & $-$ & $%
- $ \\[2pt] \hline
&  &  &  &  &  &  \\[-7pt] 
$p(\Omega_{k0}=0)$ & $43.41\,\%$ & $-$ & $85.01\,\%$ & $95.86\,\%$ & $-$ & $%
- $ \\[2pt] \hline
&  &  &  &  &  &  \\[-7pt] 
$p(\bar{A} \neq 1)$ & $0.00\,\%$ & $61.57\,\%$ & $100\,\%$ & $100\,\%$ & $%
100\,\%$ & $100\,\%$ \\[2pt] \hline
&  &  &  &  &  &  \\[-7pt] 
$p(q_{0}<0)$ & $5.20\,\sigma$ & $7.12\,\sigma$ & $100\,\%$ & $100\,\%$ & $%
100\,\%$ & $100\,\%$ \\[2pt] \hline
&  &  &  &  &  &  \\[-7pt] 
$p(a_{i}<1)$ & $5.32\,\sigma$ & $7.53\,\sigma$ & $100\,\%$ & $5.62\,\sigma$
& $100\,\%$ & $100\,\%$ \\[2pt] \hline\hline
\end{tabular}
\end{center}
\caption{The estimated parameters using the HST prior for the generalized
Chaplygin gas model (GCGM) and some specific cases of spatial section and
matter content. We use the Bayesian analysis to obtain the peak of the
one-dimensional marginal probability and the $2\,\protect\sigma $ credible
region for each parameter. $H_{0}$ is given in $km/M\!pc.s$, $\bar{A}$ in
units of $c$, $t_{0}$ in $Gy$ and $a_{i}$ in units of $a_{0}$.}
\label{tableParEstGCGp}
\end{table}

\begin{table}[t!]
\begin{center}
\begin{tabular}{|c|c|c|c|c|c|c|}
\hline\hline
&  &  &  &  &  &  \\[-7pt] 
CGM &  &  &  &  & $k=0$, & $k=0$, \\ 
with &  & $k=0$ & $\Omega _{m0}=0$ & $\Omega _{m0}=0.04$ & $\Omega _{m0}=0$
& $\Omega _{m0}=0.04$ \\[2pt] \hline
&  &  &  &  &  &  \\[-7pt] 
$H_{0}$ & $64.73_{-1.70}^{+1.72}$ & $64.48_{-1.53}^{+1.53}$ & $%
64.76_{-1.72}^{+1.74}$ & $64.74_{-1.73}^{+1.74}$ & $64.67_{-1.52}^{+1.53}$ & 
$64.64_{-1.51}^{+1.52}$ \\[2pt] \hline
&  &  &  &  &  &  \\[-7pt] 
$\Omega _{k0}$ & $-0.228_{-0.508}^{+0.552}$ & $0$ & $%
-0.099_{-0.486}^{+0.564} $ & $-0.112_{-0.481}^{+0.552}$ & $0$ & $0$ \\%
[2pt] \hline
&  &  &  &  &  &  \\[-7pt] 
$\Omega _{m0}$ & $0.000_{-0.000}^{+0.448}$ & $0.000_{-0.000}^{+0.292}$ & $0$
& $0.04$ & $0$ & $0.04$ \\[2pt] \hline
&  &  &  &  &  &  \\[-7pt] 
$\Omega _{c0}$ & $1.030_{-0.467}^{+0.448}$ & $1.000_{-0.292}^{+0.000}$ & $%
1.099_{-0.564}^{+0.486}$ & $1.072_{-0.552}^{+0.481}$ & $1$ & $0.96$ \\%
[2pt] \hline
&  &  &  &  &  &  \\[-7pt] 
$\bar{A}$ & $0.860_{-0.069}^{+0.140}$ & $0.857_{-0.072}^{+0.141}$ & $%
0.804_{-0.070}^{+0.071}$ & $0.823_{-0.068}^{+0.076}$ & $%
0.812_{-0.071}^{+0.057}$ & $0.834_{-0.072}^{+0.056}$ \\[2pt] \hline
&  &  &  &  &  &  \\[-7pt] 
$t_{0}$ & $13.93_{-0.65}^{+0.97}$ & $14.07_{-0.62}^{+0.78}$ & $%
13.95_{-0.63}^{+0.82}$ & $13.96_{-0.63}^{+0.81}$ & $13.95_{-0.60}^{+0.69}$ & 
$13.97_{-0.61}^{+0.70}$ \\[2pt] \hline
&  &  &  &  &  &  \\[-7pt] 
$q_{0}$ & $-0.769_{-0.309}^{+0.358}$ & $-0.665_{-0.108}^{+0.169}$ & $%
-0.768_{-0.339}^{+0.390}$ & $-0.780_{-0.320}^{+0.406}$ & $%
-0.708_{-0.098}^{+0.101}$ & $-0.693_{-0.090}^{+0.099}$ \\[2pt] \hline
&  &  &  &  &  &  \\[-7pt] 
$a_{i}$ & $0.689_{-0.078}^{+0.060}$ & $0.689_{-0.073}^{+0.064}$ & $%
0.702_{-0.051}^{+0.053}$ & $0.705_{-0.061}^{+0.051}$ & $%
0.704_{-0.047}^{+0.053}$ & $0.702_{-0.51}^{+0.53}$ \\[2pt] \hline
&  &  &  &  &  &  \\[-7pt] 
$p(\Omega _{k0}<0)$ & $78.91\,\%$ & $-$ & $61.83\,\%$ & $64.12\,\%$ & $-$ & $%
-$ \\[2pt] \hline
&  &  &  &  &  &  \\[-7pt] 
$p(\Omega _{k0} = 0)$ & $39.84\,\%$ & $-$ & $70.90\,\%$ & $66.80\,\%$ & $-$
& $-$ \\[2pt] \hline
&  &  &  &  &  &  \\[-7pt] 
$p(\bar{A}\neq 1)$ & $88.96\,\%$ & $96.00\,\%$ & $100\,\% $ & $3.67\,\sigma $
& $100\,\%$ & $100\,\%$ \\[2pt] \hline
&  &  &  &  &  &  \\[-7pt] 
$p(q_{0}<0)$ & $6.13\,\sigma $ & $100\,\%$ & $100\,\%$ & $100\,\%$ & $%
100\,\% $ & $100\,\%$ \\[2pt] \hline
&  &  &  &  &  &  \\[-7pt] 
$p(a_{i}<1)$ & $6.10\,\sigma $ & $100\,\%$ & $100\,\%$ & $100\,\%$ & $%
100\,\% $ & $100\,\%$ \\[2pt] \hline\hline
\end{tabular}
\end{center}
\caption{The estimated parameters for the traditional Chaplygin gas model
(CGM) and some specific cases of spatial section and matter content. We use
the Bayesian analysis to obtain the peak of the one-dimensional marginal
probability and the $2\,\protect\sigma $ credible region for each parameter. 
$H_{0}$ is given in $km/M\!pc.s$, $\bar{A}$ in units of $c$, $t_{0}$ in $Gy$
and $a_{i}$ in units of $a_{0}$. }
\label{tableParEstCG}
\end{table}

\begin{table}[t!]
\begin{center}
\begin{tabular}{|c|c|c|c|c|c|c|}
\hline\hline
&  &  &  &  &  &  \\[-7pt] 
CGM &  &  &  &  & $k=0$, & $k=0$, \\ 
with &  & $k=0$ & $\Omega _{m0}=0$ & $\Omega _{m0}=0.04$ & $\Omega _{m0}=0$
& $\Omega _{m0}=0.04$ \\[2pt] \hline
&  &  &  &  &  &  \\[-7pt] 
$H_{0}$ & $64.83_{-1.72}^{+1.69}$ & $64.53_{-1.51}^{+1.53}$ & $%
64.84_{-1.72}^{+1.72}$ & $64.83_{-1.72}^{+1.73}$ & $64.74_{-1.52}^{+1.52}$ & 
$64.70_{-1.51}^{+1.51}$ \\[2pt] \hline
&  &  &  &  &  &  \\[-7pt] 
$\Omega _{k0}$ & $-0.238_{-0.504}^{+0.548}$ & $0$ & $%
-0.109_{-0.483}^{+0.558} $ & $-0.124_{-0.477}^{+0.547}$ & $0$ & $0$ \\%
[2pt] \hline
&  &  &  &  &  &  \\[-7pt] 
$\Omega _{m0}$ & $0.000_{-0.000}^{+0.448}$ & $0.000_{-0.000}^{+0.289}$ & $0$
& $0.04$ & $0$ & $0.04$ \\[2pt] \hline
&  &  &  &  &  &  \\[-7pt] 
$\Omega _{c0}$ & $1.040_{-0.463}^{+0.446}$ & $1.000_{-0.289}^{+0.000}$ & $%
1.109_{-0.558}^{+0.483}$ & $1.084_{-0.547}^{+0.477}$ & $1$ & $0.96$ \\%
[2pt] \hline
&  &  &  &  &  &  \\[-7pt] 
$\bar{A}$ & $0.860_{-0.068}^{+0.140}$ & $0.858_{-0.072}^{+0.139}$ & $%
0.806_{-0.068}^{+0.069}$ & $0.824_{-0.066}^{+0.074}$ & $%
0.814_{-0.071}^{+0.056}$ & $0.836_{-0.070}^{+0.056}$ \\[2pt] \hline
&  &  &  &  &  &  \\[-7pt] 
$t_{0}$ & $13.93_{-0.67}^{+0.99}$ & $14.08_{-0.63}^{+0.78}$ & $%
13.95_{-0.64}^{+0.81}$ & $13.96_{-0.63}^{+0.81}$ & $13.94_{-0.58}^{+0.70}$ & 
$13.96_{-0.60}^{+0.71}$ \\[2pt] \hline
&  &  &  &  &  &  \\[-7pt] 
$q_{0}$ & $-0.780_{-0.305}^{+0.357}$ & $-0.665_{-0.111}^{+0.164}$ & $%
-0.778_{-0.336}^{+0.387}$ & $-0.789_{-0.318}^{+0.400}$ & $%
-0.711_{-0.097}^{+0.101}$ & $-0.697_{-0.088}^{+0.100}$ \\[2pt] \hline
&  &  &  &  &  &  \\[-7pt] 
$a_{i}$ & $0.687_{-0.077}^{+0.060}$ & $0.687_{-0.072}^{+0.064}$ & $%
0.700_{-0.050}^{+0.053}$ & $0.704_{-0.060}^{+0.051}$ & $%
0.702_{-0.046}^{+0.053}$ & $0.700_{-0.051}^{+0.053}$ \\[2pt] \hline
&  &  &  &  &  &  \\[-7pt] 
$p(\Omega _{k0}<0)$ & $80.14\,\% $ & $-$ & $63.57\,\%$ & $65.87\,\%$ & $-$ & 
$-$ \\[2pt] \hline
&  &  &  &  &  &  \\[-7pt] 
$p(\Omega _{k0} = 0)$ & $37.48\,\%$ & $-$ & $67.57\,\%$ & $63.47\,\%$ & $-$
& $-$ \\[2pt] \hline
&  &  &  &  &  &  \\[-7pt] 
$p(\bar{A}\neq 1)$ & $89.22\,\%$ & $96.14\,\%$ & $100\,\% $ & $3.69\,\sigma $
& $100\,\%$ & $100\,\%$ \\[2pt] \hline
&  &  &  &  &  &  \\[-7pt] 
$p(q_{0}<0)$ & $6.19\,\sigma $ & $100\,\%$ & $100\,\%$ & $100\,\%$ & $%
100\,\% $ & $100\,\%$ \\[2pt] \hline
&  &  &  &  &  &  \\[-7pt] 
$p(a_{i}<1)$ & $6.15\,\sigma $ & $100\,\%$ & $100\,\%$ & $100\,\%$ & $%
100\,\% $ & $100\,\%$ \\[2pt] \hline\hline
\end{tabular}
\end{center}
\caption{The estimated parameters using the HST prior for the traditional
Chaplygin gas model (CGM) and some specific cases of spatial section and
matter content. We use the Bayesian analysis to obtain the peak of the
one-dimensional marginal probability and the $2\,\protect\sigma $ credible
region for each parameter. $H_{0}$ is given in $km/M\!pc.s$, $\bar{A}$ in
units of $c$, $t_{0}$ in $Gy$ and $a_{i}$ in units of $a_{0}$. }
\label{tableParEstCGp}
\end{table}

\begin{table}[t!]
\begin{center}
\begin{tabular}{|c|c|c|c|c|}
\hline\hline
&  &  &  &  \\[-7pt] 
& $\Lambda$CDM & $\Lambda$CDM : & $\Lambda$CDM : & $\Lambda$CDM : \\ 
&  & $k=0$ & $\Omega _{m0}=0$ & $\Omega _{m0}=0.04$ \\[2pt] \hline
&  &  &  &  \\[-7pt] 
$H_{0}$ & $64.75_{-1.68}^{+1.68}$ & $64.29_{-1.51}^{+1.53}$ & $%
63.75_{-1.45}^{+1.58}$ & $63.86_{-1.55}^{+1.61}$ \\[2pt] \hline
&  &  &  &  \\[-7pt] 
$\Omega _{k0}$ & $-0.439_{-0.499}^{+0.636}$ & $0$ & $0.802_{-0.168}^{+0.188}$
& $0.729_{-0.159}^{+0.180}$ \\[2pt] \hline
&  &  &  &  \\[-7pt] 
$\Omega _{m0}$ & $0.459_{-0.230}^{+0.196}$ & $0.309_{-0.072}^{+0.082}$ & $0$
& $0.04$ \\[2pt] \hline
&  &  &  &  \\[-7pt] 
$\Omega _{\Lambda}$ & $0.976_{-0.426}^{+0.333}$ & $0.691_{-0.082}^{+0.072}$
& $0.198_{-0.188}^{+0.168}$ & $0.231_{-0.180}^{+0.159}$ \\[2pt] \hline
&  &  &  &  \\[-7pt] 
$t_{0}$ & $14.81_{-0.76}^{+0.87}$ & $14.80_{-0.77}^{+0.92}$ & $%
16.79_{-0.89}^{+1.16}$ & $16.03_{-0.82}^{+0.98}$ \\[2pt] \hline
&  &  &  &  \\[-7pt] 
$q_{0}$ & $-0.746_{-0.271}^{+0.332}$ & $-0.532_{-0.116}^{+0.121}$ & $%
-0.187_{-0.183}^{+0.170}$ & $-0.213_{-0.158}^{+0.183}$ \\[2pt] \hline
&  &  &  &  \\[-7pt] 
$a_{i}$ & $0.620_{-0.059}^{+0.060}$ & $0.604_{-0.070}^{+0.083}$ & $0$ & $%
0.448_{-0.159}^{+0.204}$ \\[2pt] \hline
&  &  &  &  \\[-7pt] 
$p(\Omega _{k0}<0)$ & $90.60\,\%$ & $-$ & $0\,\%$ & $0\,\%$ \\[2pt] \hline
&  &  &  &  \\[-7pt] 
$p(\Omega _{k0} = 0)$ & $15.44\,\%$ & $-$ & $0\,\%$ & $0\,\%$ \\[2pt] \hline
&  &  &  &  \\[-7pt] 
$p(q_{0}<0)$ & $7.24\,\sigma $ & $100\,\%$ & $100\,\% $ & $99.26\,\% $ \\%
[2pt] \hline
&  &  &  &  \\[-7pt] 
$p(a_{i}<1)$ & $7.24\,\sigma $ & $100\,\%$ & $100\,\% $ & $99.38\,\% $ \\%
[2pt] \hline\hline
\end{tabular}
\end{center}
\caption{The estimated parameters for the $\Lambda$CDM model and some
specific cases of spatial section and matter content. We use the Bayesian
analysis to obtain the peak of the one-dimensional marginal probability and
the $2\,\protect\sigma $ credible region for each parameter. $H_{0}$ is
given in $km/M\!pc.s$, $t_{0}$ in $Gy$ and $a_{i}$ in units of $a_{0}$. }
\label{tableParEstLCDM}
\end{table}

\begin{table}[t!]
\begin{center}
\begin{tabular}{|c|c|c|c|c|}
\hline\hline
&  &  &  &  \\[-7pt] 
& $\Lambda$CDM & $\Lambda$CDM : & $\Lambda$CDM : & $\Lambda$CDM : \\ 
&  & $k=0$ & $\Omega _{m0}=0$ & $\Omega _{m0}=0.04$ \\[2pt] \hline
&  &  &  &  \\[-7pt] 
$H_{0}$ & $64.82_{-1.66}^{+1.68}$ & $64.36_{-1.50}^{+1.52}$ & $%
63.83_{-1.46}^{+1.58}$ & $63.94_{-1.55}^{+1.60}$ \\[2pt] \hline
&  &  &  &  \\[-7pt] 
$\Omega _{k0}$ & $-0.449_{-0.496}^{+0.631}$ & $0$ & $0.794_{-0.170}^{+0.191}$
& $0.683_{-0.185}^{+0.210}$ \\[2pt] \hline
&  &  &  &  \\[-7pt] 
$\Omega _{m0}$ & $0.460_{-0.228}^{+0.195}$ & $0.306_{-0.071}^{+0.081}$ & $0$
& $0.04$ \\[2pt] \hline
&  &  &  &  \\[-7pt] 
$\Omega _{\Lambda}$ & $0.985_{-0.423}^{+0.330}$ & $0.694_{-0.081}^{+0.071}$
& $0.206_{-0.191}^{+0.170}$ & $0.277_{-0.210}^{+0.185}$ \\[2pt] \hline
&  &  &  &  \\[-7pt] 
$t_{0}$ & $14.83_{-0.77}^{+0.86}$ & $14.83_{-0.78}^{+0.92}$ & $%
16.82_{-0.92}^{+1.16}$ & $16.06_{-0.83}^{+0.98}$ \\[2pt] \hline
&  &  &  &  \\[-7pt] 
$q_{0}$ & $-0.755_{-0.267}^{+0.330}$ & $-0.537_{-0.115}^{+0.121}$ & $%
-0.197_{-0.182}^{+0.176}$ & $-0.256_{-0.186}^{+0.211}$ \\[2pt] \hline
&  &  &  &  \\[-7pt] 
$a_{i}$ & $0.620_{-0.060}^{+0.058}$ & $0.604_{-0.072}^{+0.081}$ & $0$ & $%
0.380_{-0.095}^{+0.215}$ \\[2pt] \hline
&  &  &  &  \\[-7pt] 
$p(\Omega _{k0}<0)$ & $91.28\,\%$ & $-$ & $0\,\%$ & $0\,\%$ \\[2pt] \hline
&  &  &  &  \\[-7pt] 
$p(\Omega _{k0} = 0)$ & $14.27\,\%$ & $-$ & $0\,\%$ & $0\,\%$ \\[2pt] \hline
&  &  &  &  \\[-7pt] 
$p(q_{0}<0)$ & $7.29\,\sigma $ & $100\,\%$ & $100\,\% $ & $99.50\,\% $ \\%
[2pt] \hline
&  &  &  &  \\[-7pt] 
$p(a_{i}<1)$ & $7.29\,\sigma $ & $100\,\%$ & $100\,\% $ & $99.57\,\% $ \\%
[2pt] \hline\hline
\end{tabular}
\end{center}
\caption{The estimated parameters using the HST prior for the $\Lambda$CDM
model and some specific cases of spatial section and matter content. We use
the Bayesian analysis to obtain the peak of the one-dimensional marginal
probability and the $2\,\protect\sigma $ credible region for each parameter. 
$H_{0}$ is given in $km/M\!pc.s$, $t_{0}$ in $Gy$ and $a_{i}$ in units of $%
a_{0}$. }
\label{tableParEstLCDMp}
\end{table}

\begin{figure}[ht!]
\begin{minipage}[t]{0.47\linewidth}
\includegraphics[trim=0.4in 0.2in 0.2in 0in,width=\linewidth]{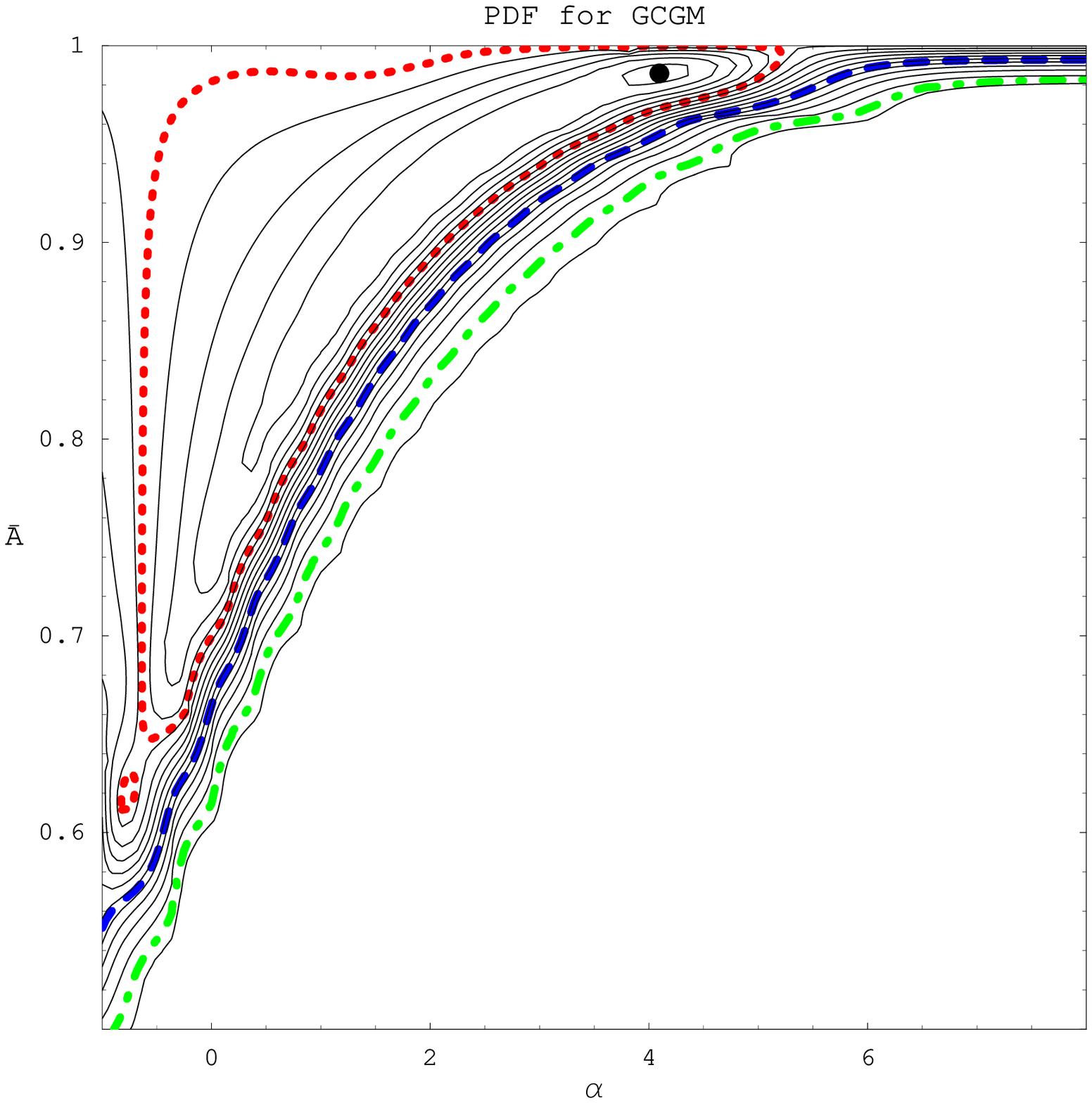}
\end{minipage} \hfill 
\begin{minipage}[t]{0.47\linewidth}
\includegraphics[trim=0.2in 0.2in 0.4in 0in,width=\linewidth]{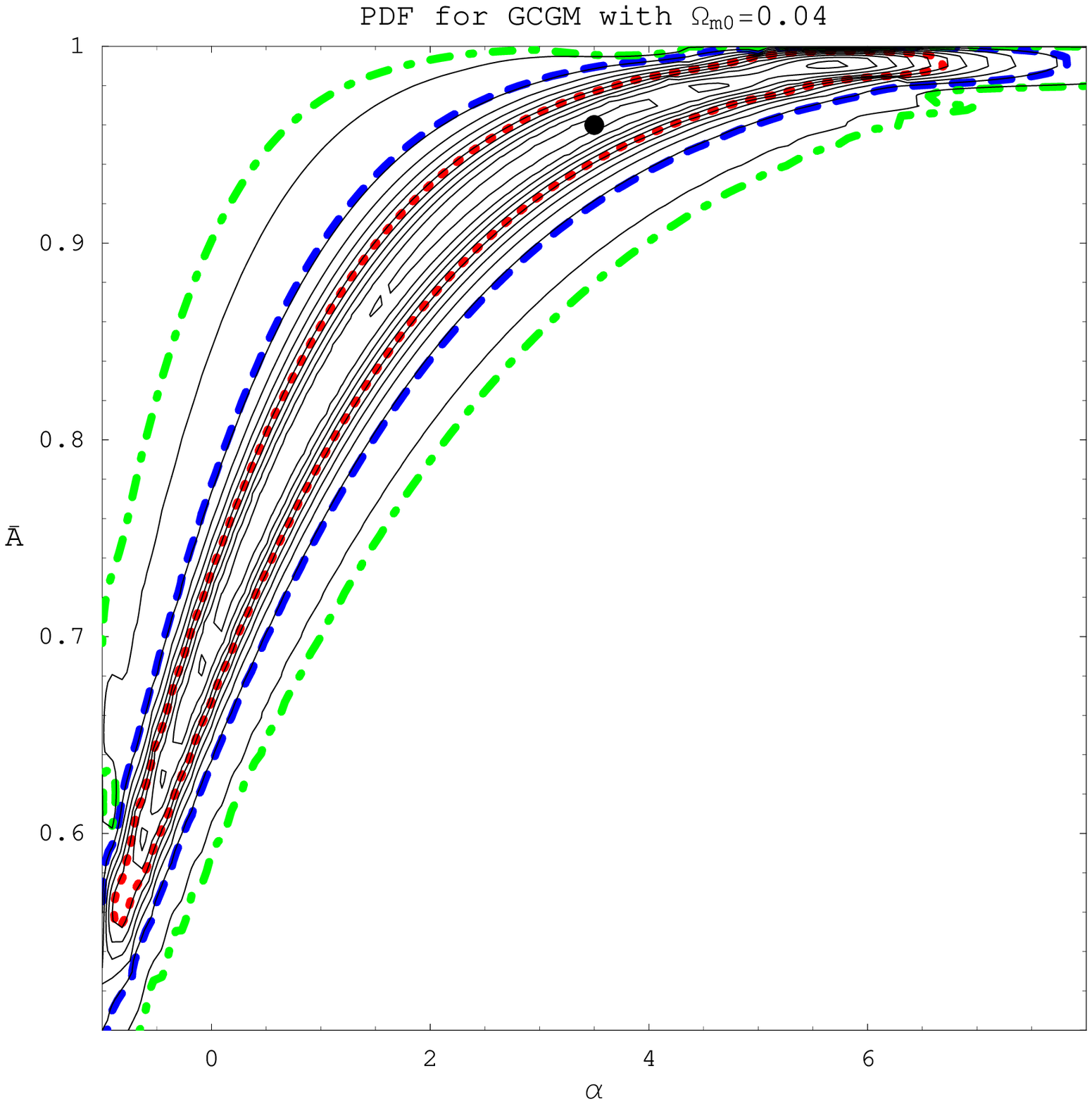}
\end{minipage} \hfill
\caption{{\protect\footnotesize The graphics of the joint PDF as function of 
$(\protect\alpha,\bar{A})$ for the generalized Chaplygin gas model. The
joint PDF peak is shown by the large dot, the credible regions of $1\,%
\protect\sigma $ ($68,27\%$) by the red dotted line, the $2\,\protect\sigma $
($95,45\%$) in blue dashed line and the $3\,\protect\sigma $ ($99,73\%$) in
green dashed-dotted line. The cases for $\Omega _{m0}=0$ are not shown here
because they are similar to the ones with $\Omega _{m0}=0.04$. The cases for 
$k=0$ are shown in figure $2$ of ref. \protect\cite{colistete3}. }}
\label{figsAlphaA}
\end{figure}

\begin{figure}[ht!]
\begin{minipage}[t]{0.48\linewidth}
\includegraphics[trim=0.4in 0.2in 0.2in 0in,width=\linewidth]{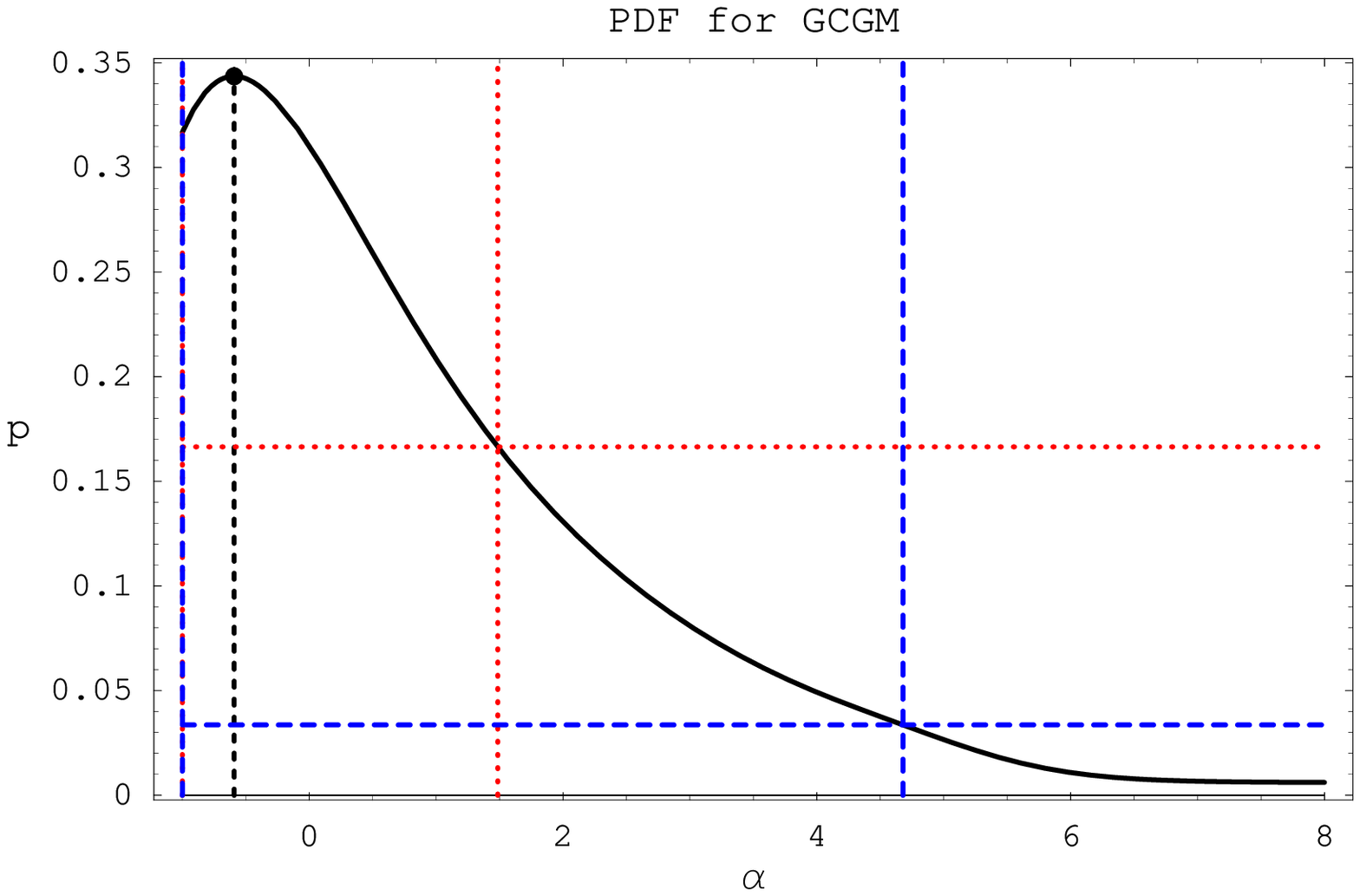}
\end{minipage} \hfill 
\begin{minipage}[t]{0.48\linewidth}
\includegraphics[trim=0.2in 0.2in 0.4in 0in,width=\linewidth]{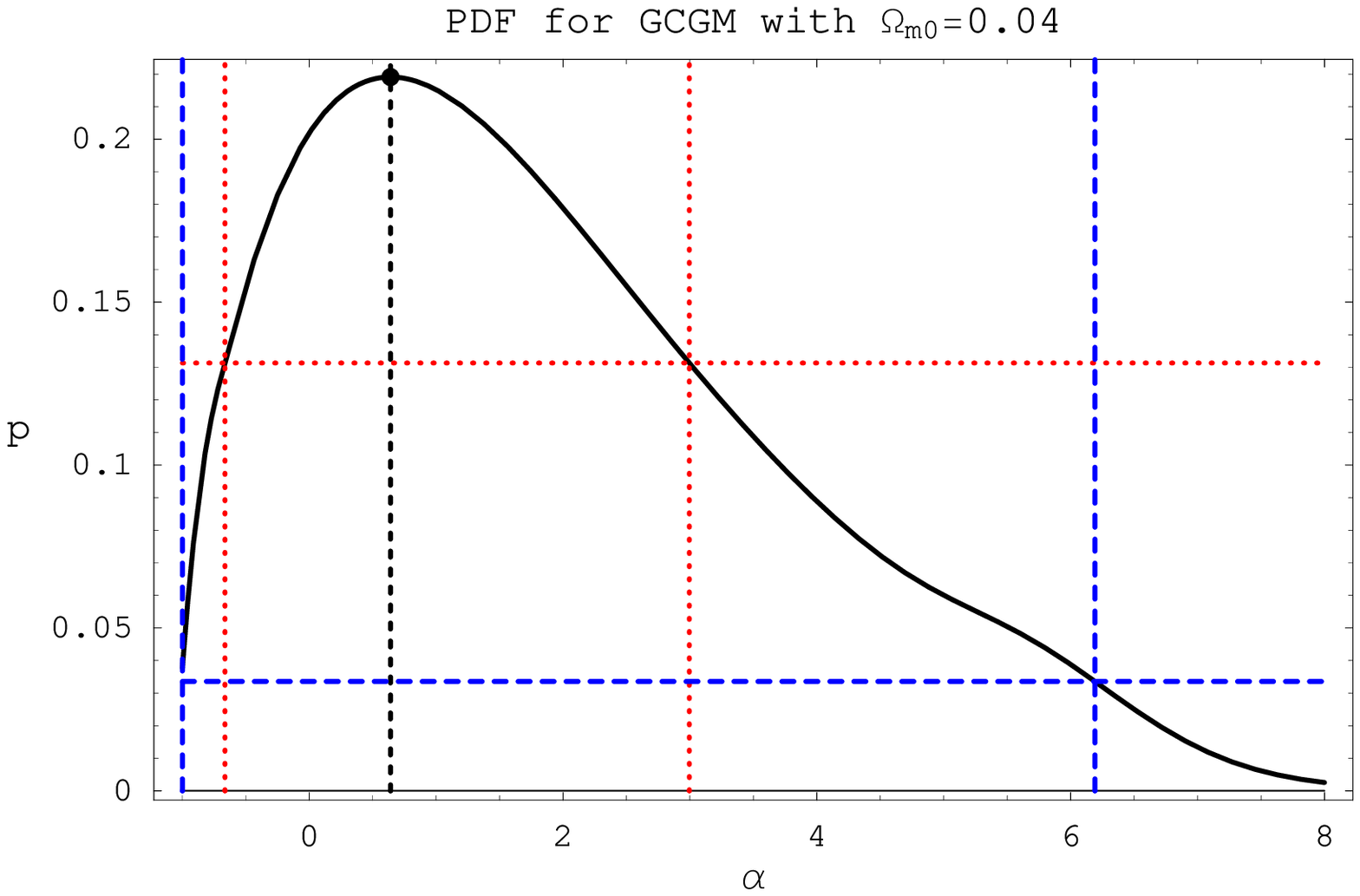}
\end{minipage} \hfill
\caption{{\protect\footnotesize The PDF of $\protect\alpha$ for the
generalized Chaplygin gas model. The solid lines are the PDF, the $1\protect%
\sigma $ ($68.27\%$) regions are delimited by red dotted lines and the $2 
\protect\sigma $ ($95.45\%$) credible regions are given by blue dashed
lines. The cases for $\Omega _{m0}=0$ are not shown here because they are
similar to the ones with $\Omega _{m0}=0.04$. The cases for $k=0$ are shown
in figure $3$ of ref. \protect\cite{colistete3}.}}
\label{figsAlpha}
\end{figure}

Note that the marginalized estimations differs substantially from those
extracted from the minimization of $\chi^{2}$, which gives a large positive
best value $\alpha$, but the dispersion is quite high, so even large
positive values are not excluded, at least at $2\sigma $ level. And figure 
\ref{figsAlphaA} for the joint probabilities for $\alpha$ and $\bar A$ and
figures \ref{figsAlpha} and \ref{figsA} for $\alpha$ and $\bar A$,
respectively, clearly show that the marginalization process changes the peak
values and credible regions depending on the number of dimensions.

Of course, the CGM is obtained when $\alpha$ is fixed to unity. From the
analysis of the GCGM it can be inferred that $p(\alpha = 1)=40.81\%$, i.e.,
the CGM is favoured with a probability of $40.81\%$. Restricting to null
curvature or fixing the pressureless matter density increases considerably
this value, from $68.22\%$ to $91.09\%$. Analogously, the probability to
have $\alpha > 0$ (with more physical meaning) is $66.74\%$ and this value
is quite increased when one or two parameters are fixed. Both $p(\alpha = 1)$
and $p(\alpha > 0)$ are increased with respect to ref. \cite{colistete3}.

\subsection{Estimation of $\bar A$}

Like ref. \cite{colistete3}, the results indicate that the value of $\bar{A}$
is close to unity, but now the dispersion is slightly smaller. In the case
of GCGM with no fixed parameters, the marginalization of the remaining four
other parameters leads to $\bar{A}=1.000_{-0.348}^{+0.000}$. Again, this
could suggest the conclusion that $\Lambda $CDM ($\bar{A}=1$) model is
favoured. However, the accuracy of the computation, due to the step (between 
$0.01$ and $0.02$) used in the evaluation of the parameter, does not allow
this conclusion. Instead, it means the peak happens for $0.98<\bar{A}%
\leqslant 1$. In fact, fixing the curvature or the pressureless matter, the
preferred value differs slightly from unity, for example the quartessence
scenario, $\Omega _{m}=0$, yields $\bar{A}=0.987_{-0.3888}^{+0.012}$. But,
differently from ref. \cite{colistete3}, the CGM now predicts a best value
for $\bar{A}$ smaller than unity, $\bar{A}=0.860_{-0.069}^{+0.140}$, and the
best value becomes smaller when one or two parameters are fixed.

\begin{figure}[ht!]
\begin{minipage}[t]{0.48\linewidth}
\includegraphics[trim=0.4in -0.2in 0.2in 0in,width=\linewidth]{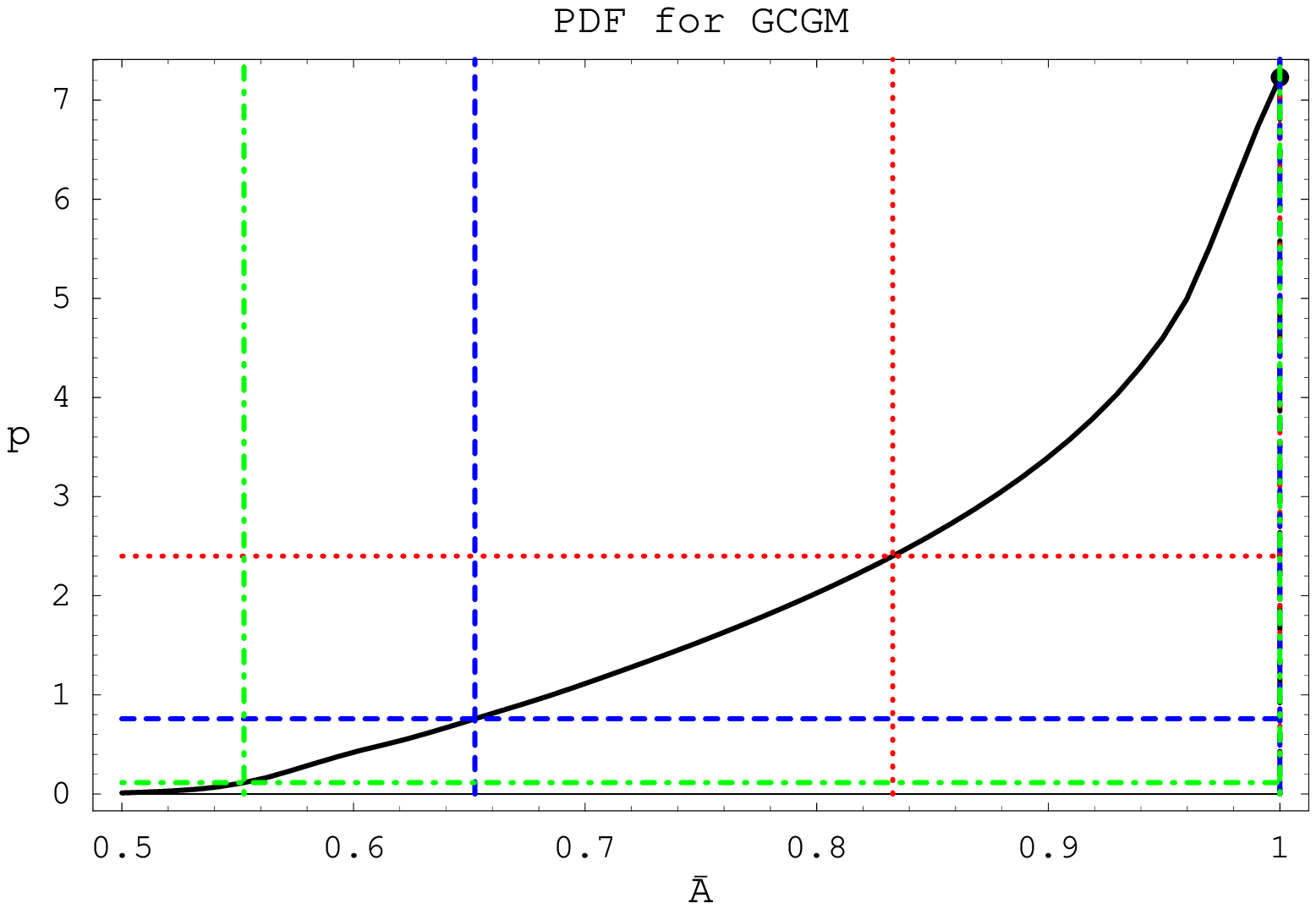}
\end{minipage} \hfill 
\begin{minipage}[t]{0.48\linewidth}
\includegraphics[trim=0.2in -0.2in 0.4in 0in,width=\linewidth]{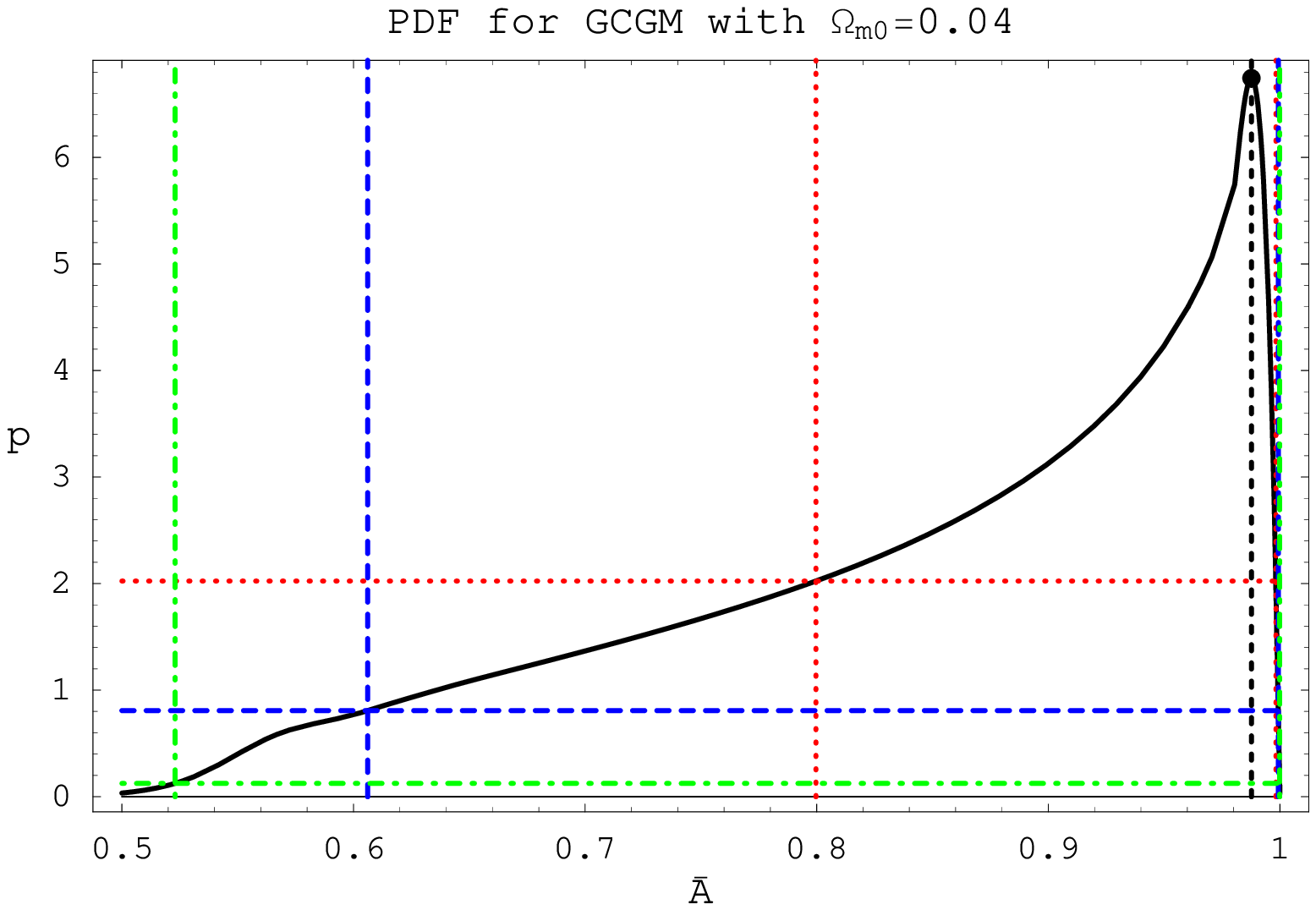}
\end{minipage} \hfill 
\begin{minipage}[t]{0.48\linewidth}
\includegraphics[trim=0.4in 0.2in 0.2in 0in,width=\linewidth]{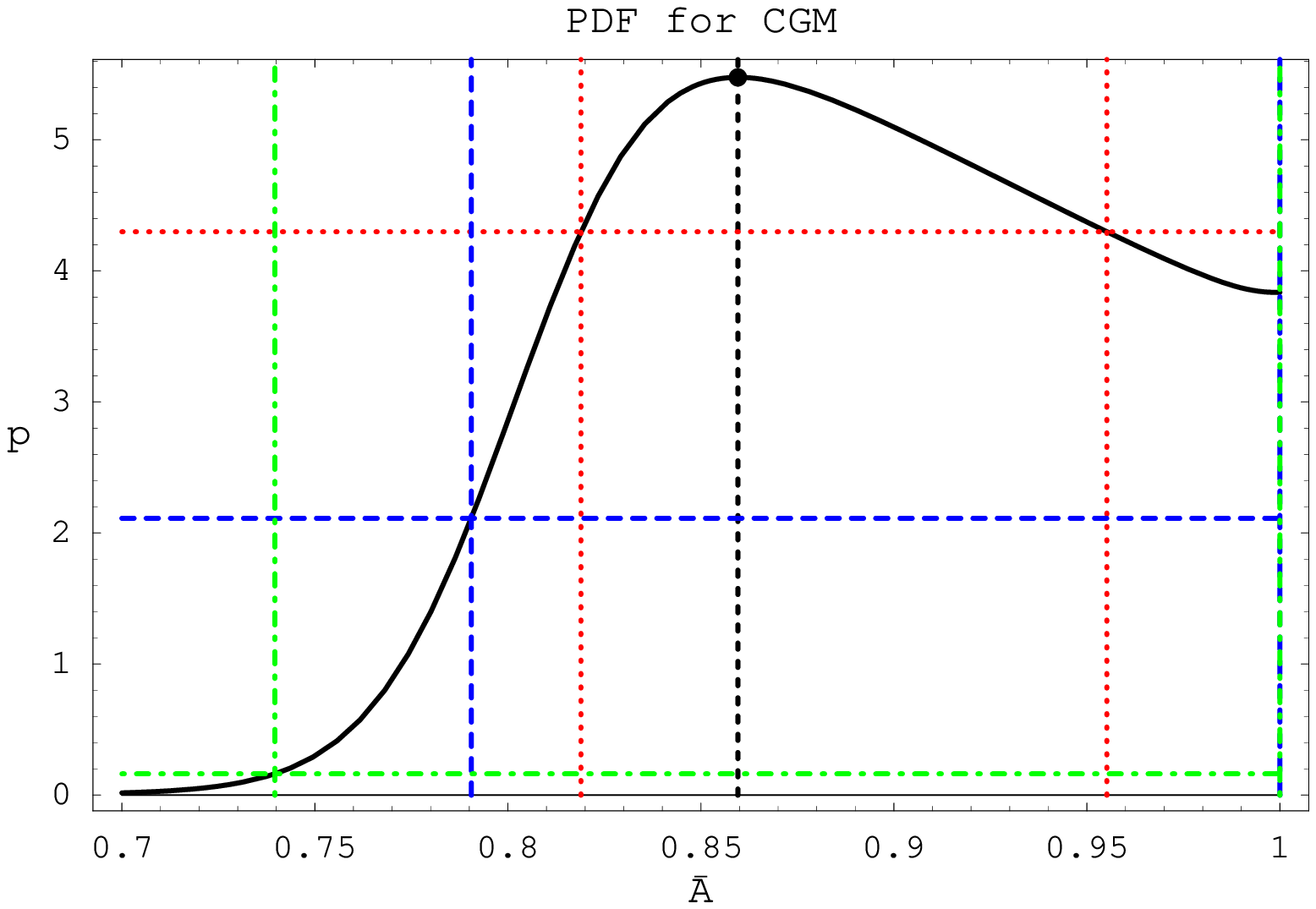}
\end{minipage} \hfill 
\begin{minipage}[t]{0.48\linewidth}
\includegraphics[trim=0.2in 0.2in 0.4in 0in,width=\linewidth]{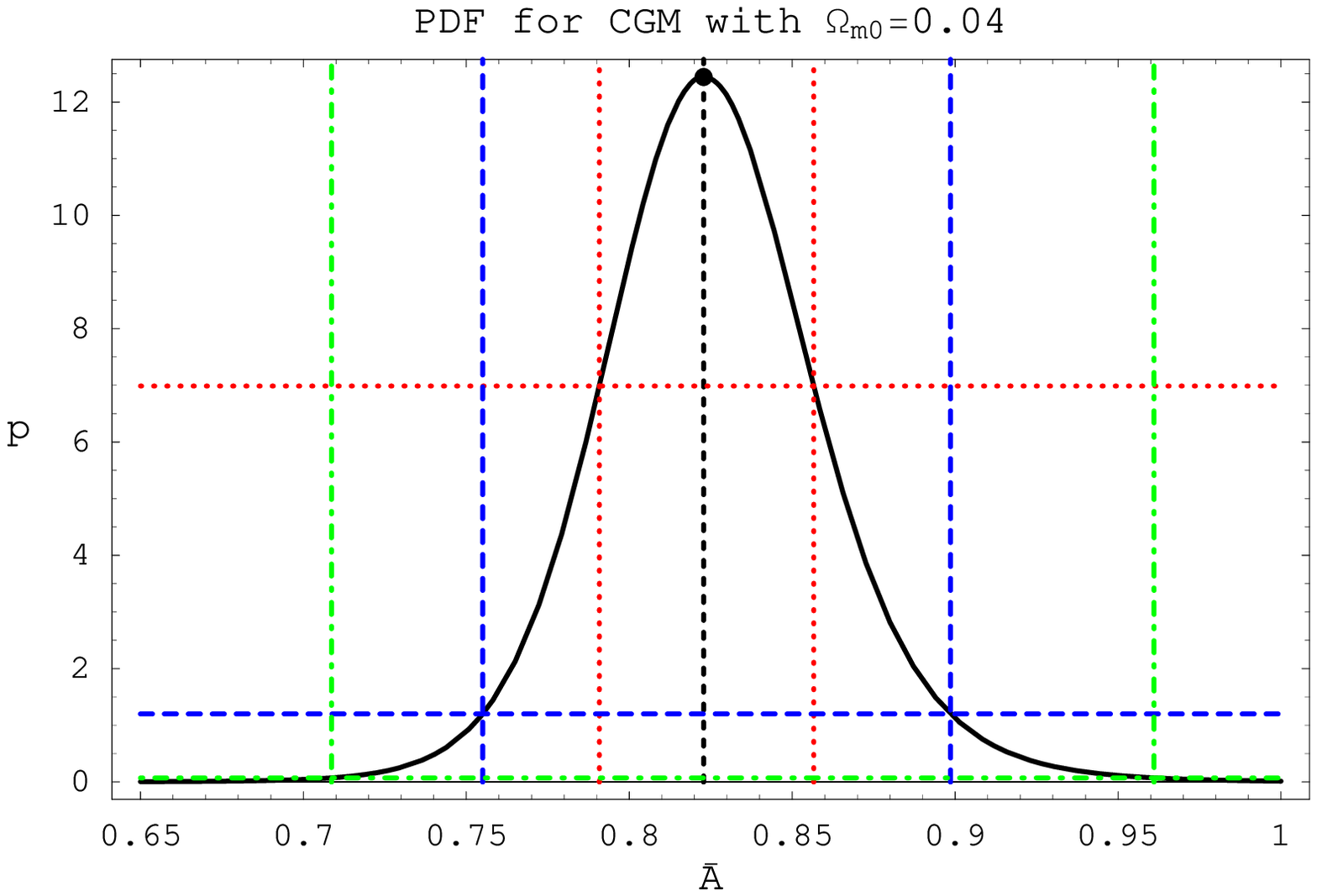}
\end{minipage} \hfill
\caption{{\protect\footnotesize The one-dimensional PDF of $\bar{A}$ for the
generalized and traditional Chaplygin gas model. The solid lines are the
PDF, the $1\protect\sigma $ ($68.27\%$) regions are delimited by red dotted
lines, the $2\protect\sigma $ ($95.45\%$) credible regions are given by blue
dashed lines and the $3\protect\sigma $ ($99.73\%$) regions are delimited by
green dashed-dotted lines. The cases for $\Omega _{m0}=0$ are not shown here
because they are similar to the ones with $\Omega _{m0}=0.04$. The cases for 
$k=0$ are shown in figures $4$ and $5$ of ref. \protect\cite{colistete3}. }}
\label{figsA}
\end{figure}

In figure \ref{figsA} the PDF for $\bar{A}$ is displayed, both for the GCGM
and the CGM, where the marginalization is made in all other parameters. Note
that the probability to have $\bar{A}\neq 1$ (meaning how much the $\Lambda$%
CDM is rule out) is zero only for the GCGM (due to the step used in the
evaluation of $\bar{A}$), but this probability varies from about $60\%$ to $%
100\%$ for other cases.

In figure \ref{figsAlphaA} the joint probabilities for $\alpha$ and $\bar A$
are displayed, with a non-Gaussian shape. Comparing with ref. \cite
{colistete3}, the peak values now happen for large values of $\alpha$ and $%
\bar A$. This figure, compared to figures \ref{figsAlpha} and \ref{figsA},
is an illustration of the importance of the marginalization process because
it changes the peak values and credible regions depending on whether two or
one-dimensional parameter space is used.

\subsection{Estimation of $\Omega_{m0}$ and $\Omega_{c0}$}

\begin{figure}[ht!]
\begin{minipage}[t]{0.45\linewidth}
\includegraphics[trim=0.5in 0.2in 0.4in 0in,width=\linewidth]{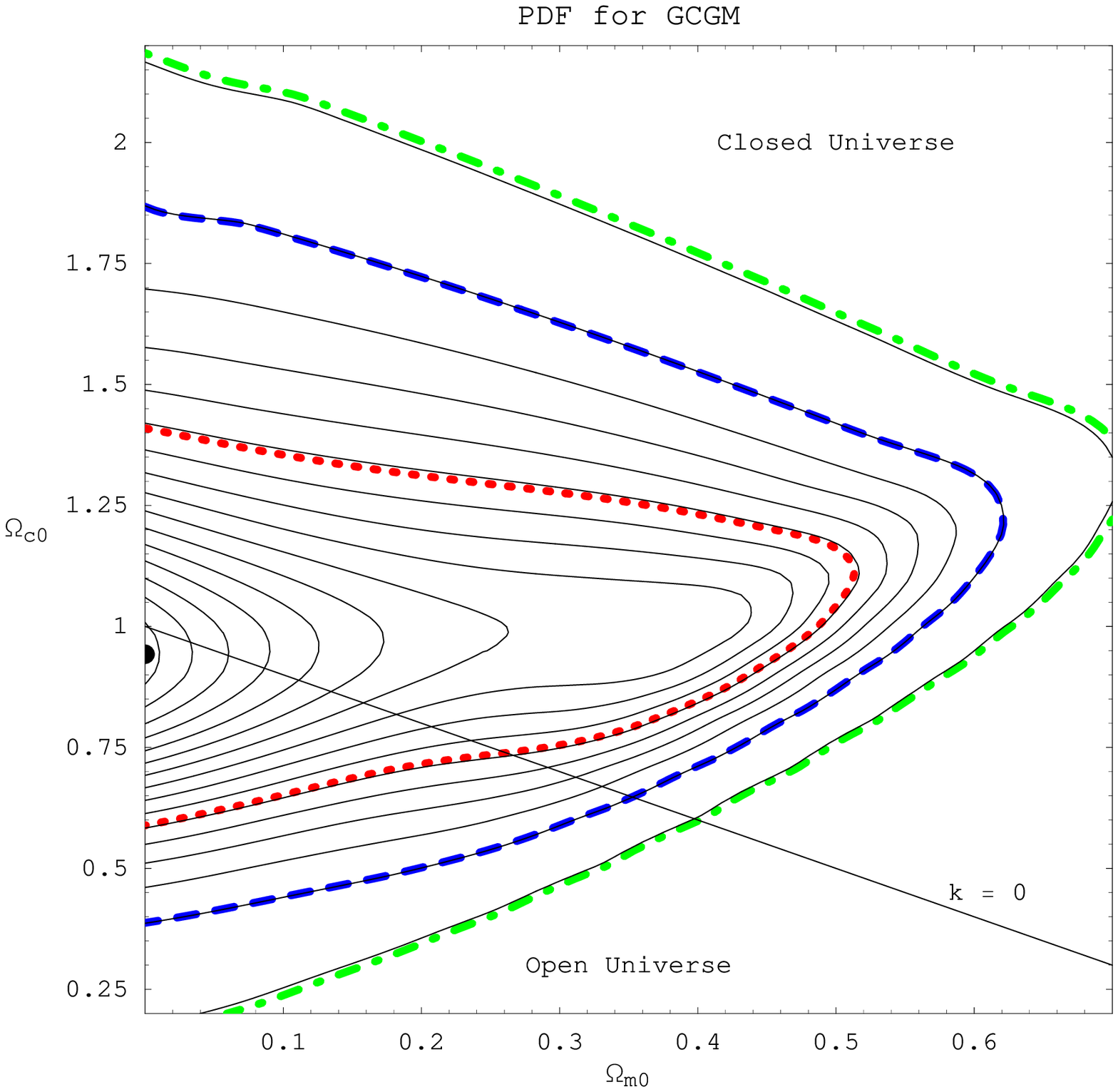}
\end{minipage} \hfill 
\begin{minipage}[t]{0.45\linewidth}
\includegraphics[trim=0.5in 0.2in 0.4in 0in,width=\linewidth]{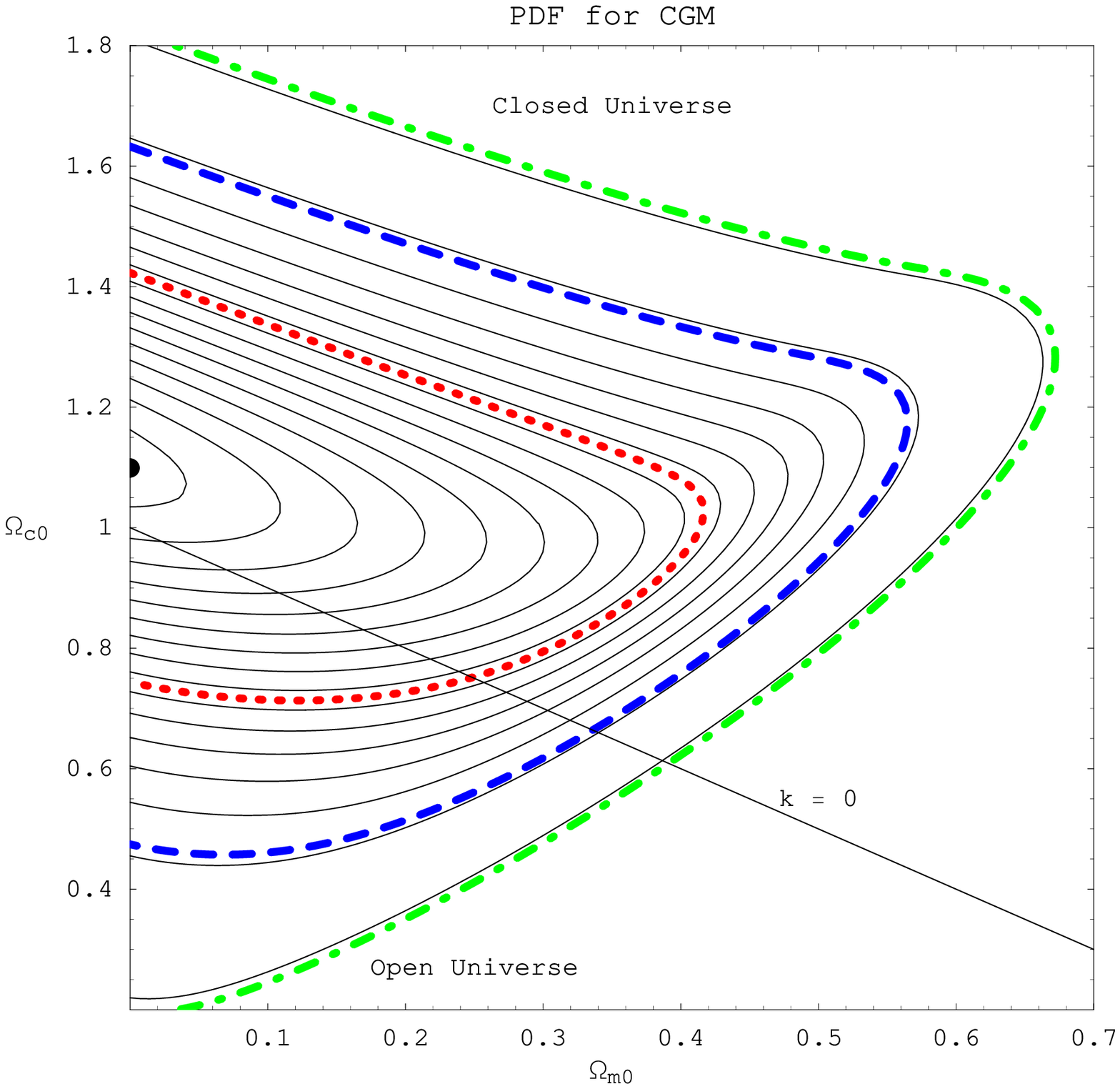}
\end{minipage} \hfill
\caption{{\protect\footnotesize The graphics of the joint PDF as function of 
$(\Omega _{m0},\Omega _{c0})$ for the generalized Chaplygin gas model, GCGM
(traditional Chaplygin gas model, CGM), where $p(\Omega _{m0},\Omega _{c0})$
is a integral of $p(\protect\alpha,H_{0},\Omega _{m0},\Omega _{c0},\bar{A})$
($p(H_{0},\Omega_{m0},\Omega _{c0},\bar{A})$) over the $(\protect\alpha
,H_{0},\bar{A})$ ($(H_{0}, \bar{A})$) parameter space. The joint normalized
PDF peak has the value $4.624$ ($5.158$) for $(\Omega
_{m0},\Omega_{c0})=(0.000,0.943)$ ($(\Omega _{m0},\Omega_{c0})=(0.000,1.099)$%
) shown by the large dot, the credible regions of $1\,\protect\sigma $ ($%
68,27\%$, shown in red dotted line), $2\,\protect\sigma $ ($95,45\%$, in
blue dashed line) and $3\,\protect\sigma $ ($99,73\%$, in green
dashed-dotted line) have PDF levels of $1.577$, $0.345$\ and $0.038$ ($2.191$%
, $0.439$\ and $0.040$), respectively. As $\Omega _{k0}+\Omega _{m0}+\Omega
_{c0}=1$, the probability for a spatially flat Universe is on the line $%
\Omega _{m0}+\Omega _{c0}=1$, above it we have the region for a closed
Universe ($k>0$, $\Omega _{k0}<0$), and below, the region for an open
Universe ($k<0$, $\Omega _{k0}>0$). }}
\label{figCGCCGOmegam0c0}
\end{figure}

\begin{figure}[ht!]
\begin{center}
\includegraphics[trim=0in 0.4in 0in
0in,scale=0.6]{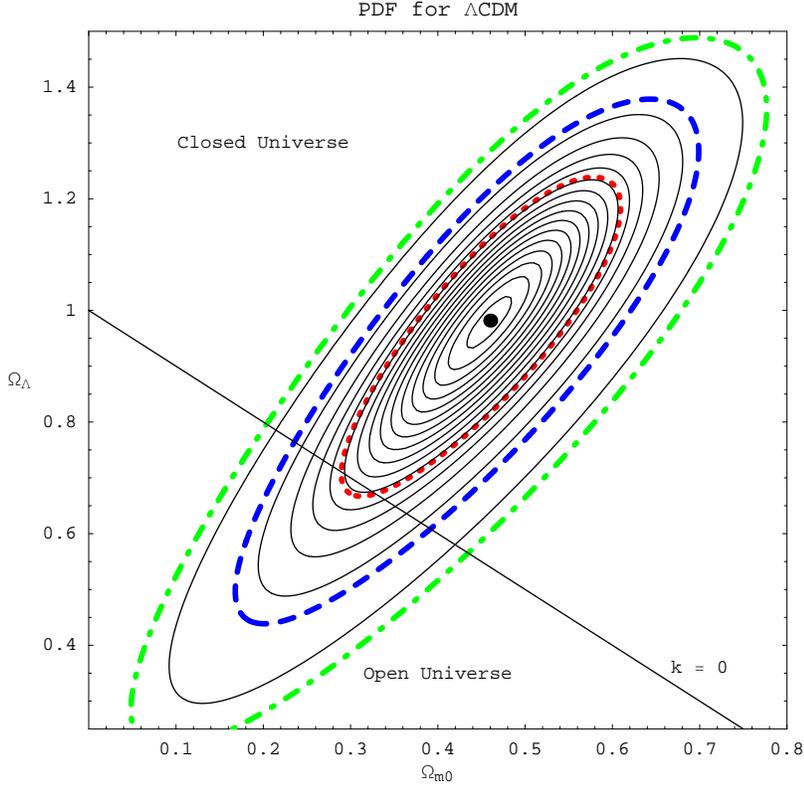}
\end{center}
\caption{{\protect\footnotesize The graphics of the joint PDF as function of 
$(\Omega _{m0},\Omega _{c0})$ for the $\Lambda$CDM model model, where $%
p(\Omega _{m0},\Omega _{c0})$ is a integral of $p(H_{0},\Omega _{m0},\Omega
_{c0})$ over the $H_{0}$ parameter space. The joint normalized PDF peak has
the value $14.71$ for $(\Omega _{m0},\Omega _{c0})=(0.460,0.981)$ (shown by
the large dot), the credible regions of $1\,\protect\sigma $ ($68,27\%$,
shown in red dotted line), $2\,\protect\sigma $ ($95,45\%$, in blue dashed
line) and $3\,\protect\sigma $ ($99,73\%$, in green dashed-dotted line) have
PDF levels of $4.634$, $0.675$\ and $0.056$, respectively. As $\Omega
_{k0}+\Omega_{m0}+\Omega _{c0}=1$, the probability for a spatially flat
Universe is on the line $\Omega _{m0}+\Omega _{c0}=1$, above it we have the
region for a closed Universe ($k>0$, $\Omega _{k0}<0$), and below, the
region for an open Universe ($k<0$, $\Omega _{k0}>0$). }}
\label{figLCDMOmegam0c0}
\end{figure}

\begin{figure}[ht!]
\begin{minipage}[t]{0.48\linewidth}
\includegraphics[trim=0.4in -0.2in 0.2in 0in,width=\linewidth]{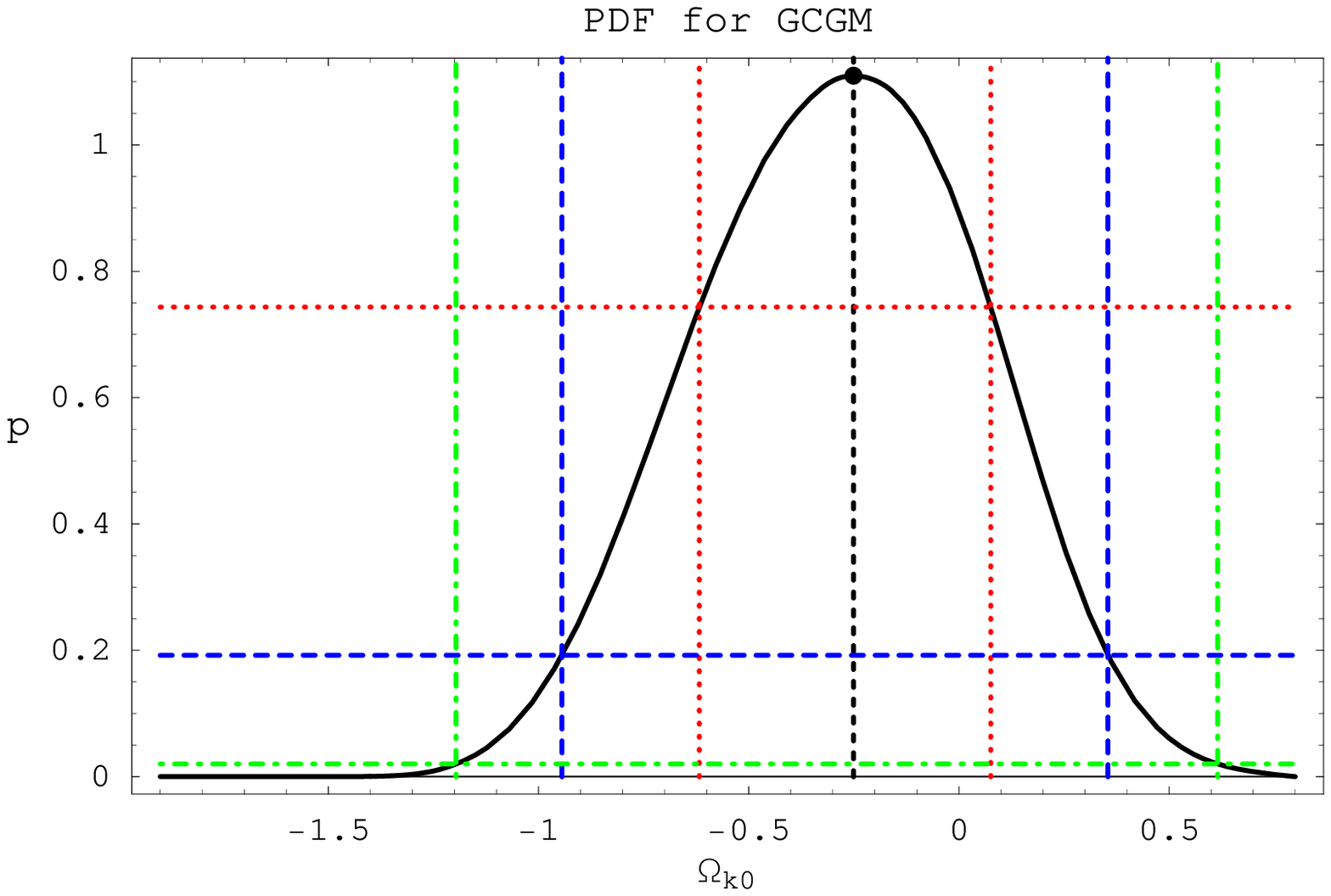}
\end{minipage} \hfill 
\begin{minipage}[t]{0.48\linewidth}
\includegraphics[trim=0.2in -0.2in 0.4in 0in,width=\linewidth]{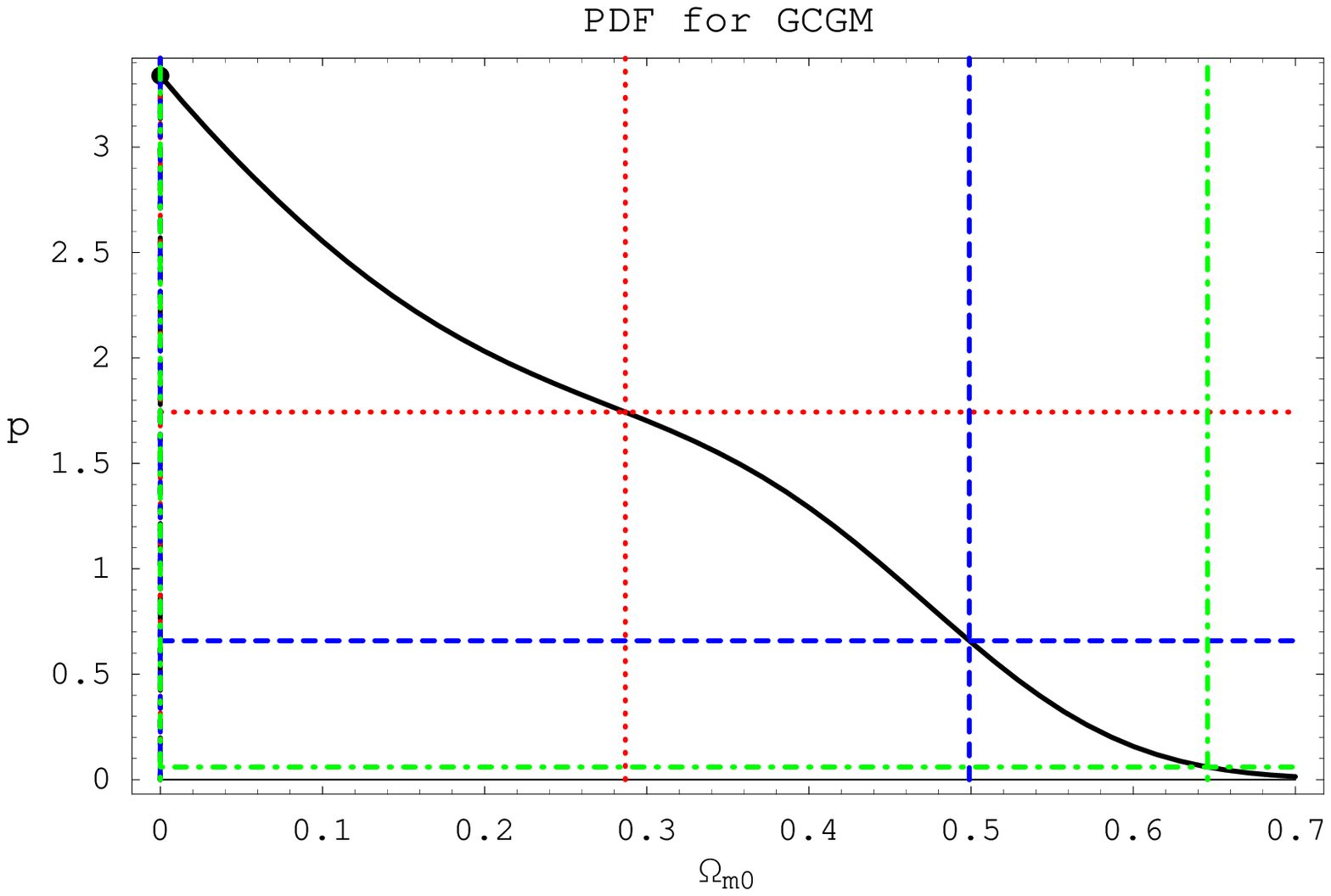}
\end{minipage} \hfill 
\begin{minipage}[t]{0.48\linewidth}
\includegraphics[trim=0.4in 0.2in 0.2in 0in,width=\linewidth]{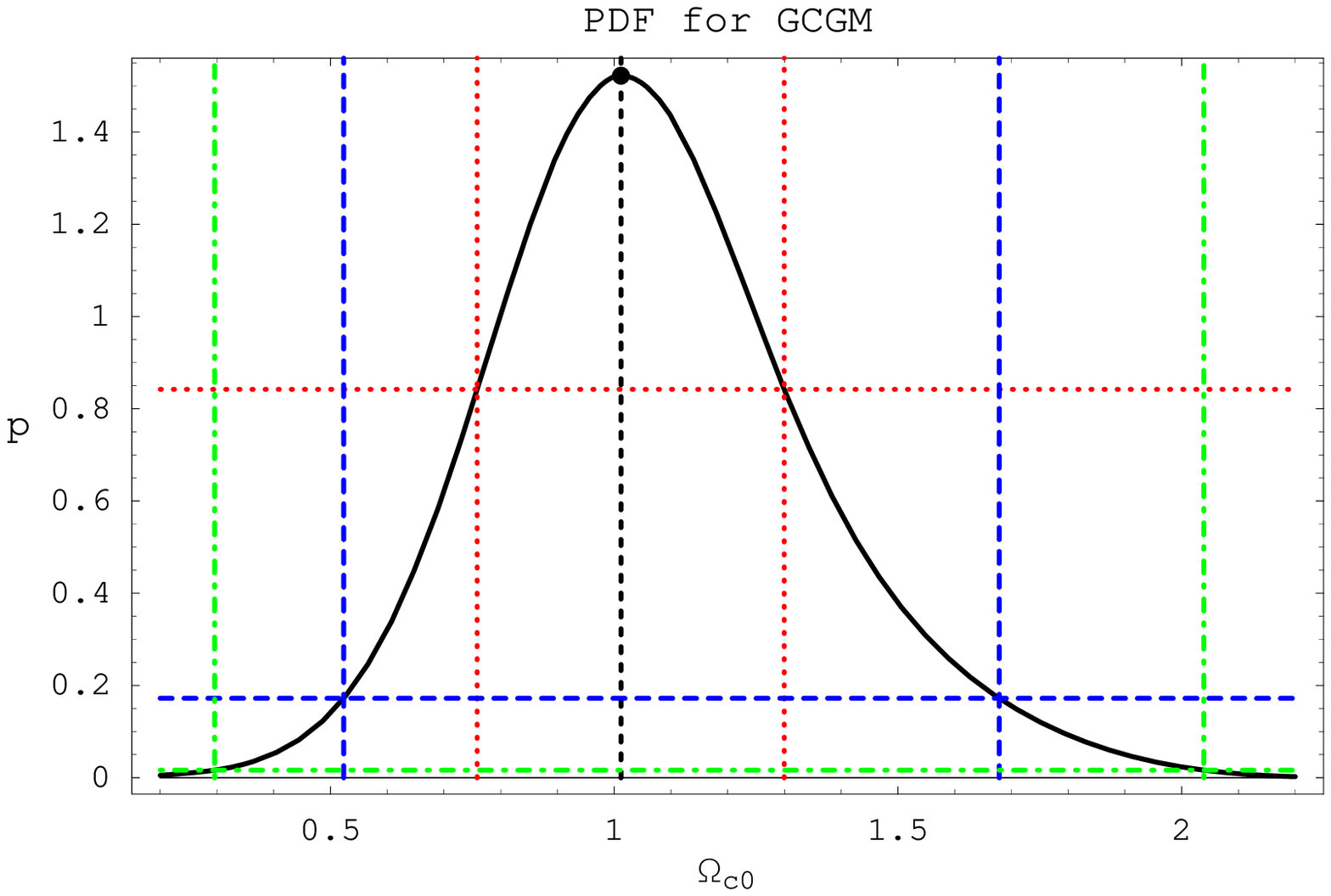}
\end{minipage} \hfill 
\begin{minipage}[t]{0.48\linewidth}
\includegraphics[trim=0.2in 0.2in 0.4in 0in,width=\linewidth]{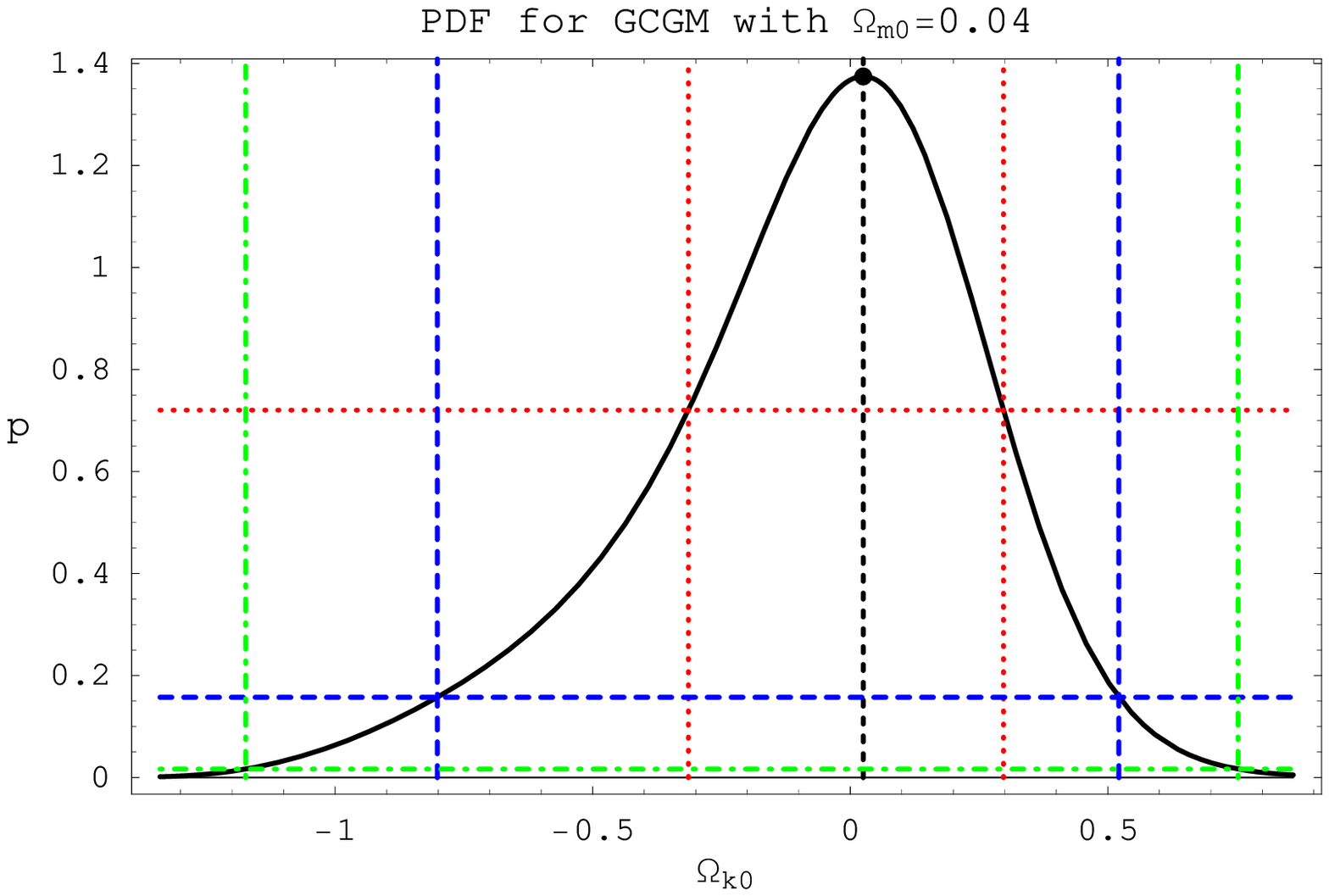}
\end{minipage} \hfill
\caption{{\protect\footnotesize The one-dimensional PDF of $\Omega _{k0}$, $%
\Omega _{m0}$ and $\Omega _{c0}$ for the generalized Chaplygin gas model.
The solid lines are the PDF, the $1\protect\sigma $ ($68.27\%$) regions are
delimited by red dotted lines, the $2\protect\sigma $ ($95.45\%$) credible
regions are given by blue dashed lines and the $3\protect\sigma $ ($99.73\%$%
) regions are delimited by green dashed-dotted lines. The case for $%
\Omega_{m0}=0$ is not shown here because it is similar to the one with $%
\Omega_{m0}=0.04$. The case for $k=0$ is shown in figure $9$ of ref. 
\protect\cite{colistete3}. As $\Omega _{c0}=1-\Omega_{k0}-\Omega _{k0}$, for 
$\Omega _{m0}=0$ we have $\Omega _{c0}=1-\Omega_{k0}$, for $\Omega
_{m0}=0.04 $ then $\Omega _{c0}=0.96-\Omega _{k0}$ and for $\Omega _{k0}=0$
we also have $\Omega _{c0}=1-\Omega _{m0}$. }}
\label{figsGCGOmegas}
\end{figure}

\begin{figure}[ht!]
\begin{minipage}[t]{0.48\linewidth}
\includegraphics[trim=0.4in -0.2in 0.2in 0in,width=\linewidth]{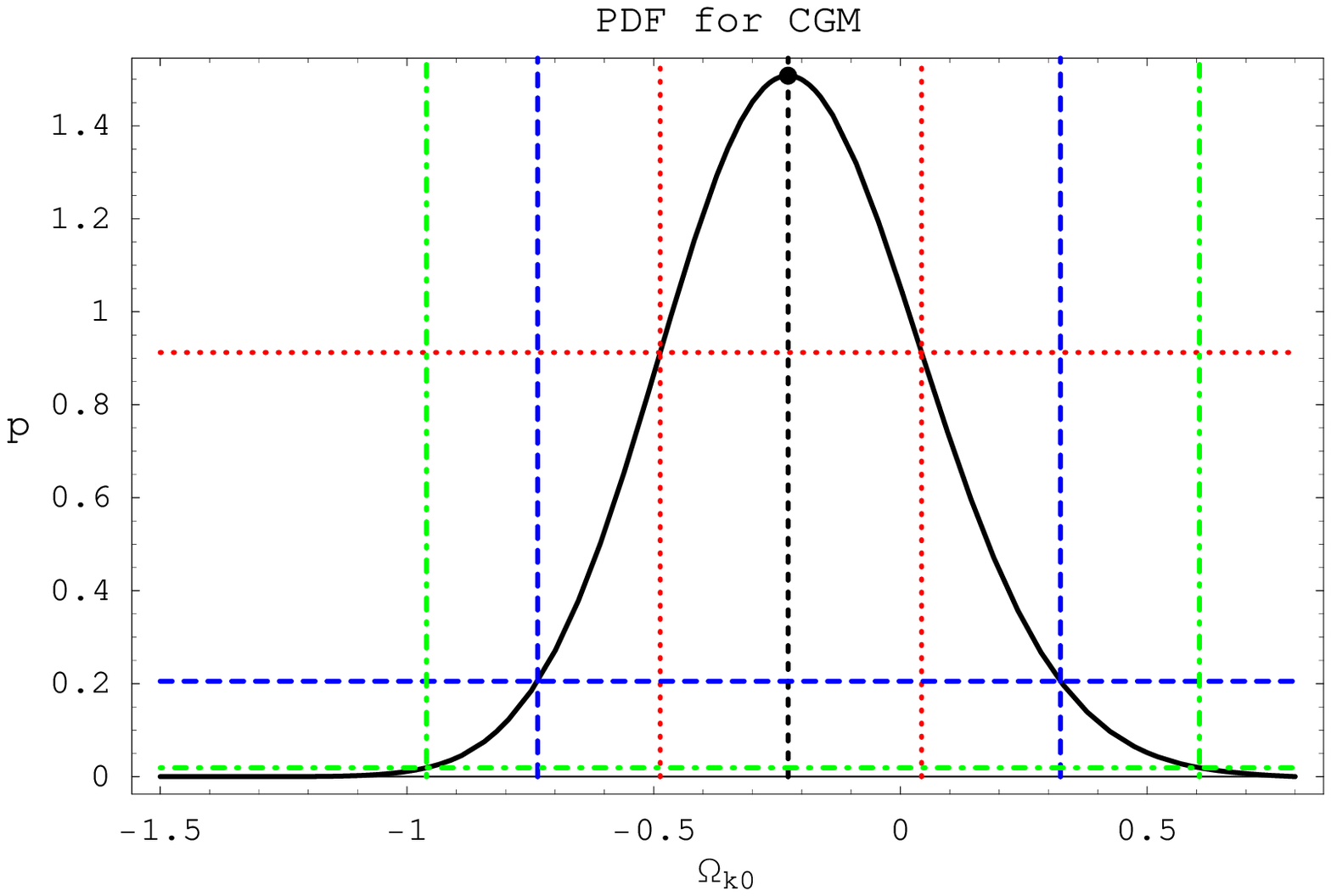}
\end{minipage} \hfill 
\begin{minipage}[t]{0.48\linewidth}
\includegraphics[trim=0.2in -0.2in 0.4in 0in,width=\linewidth]{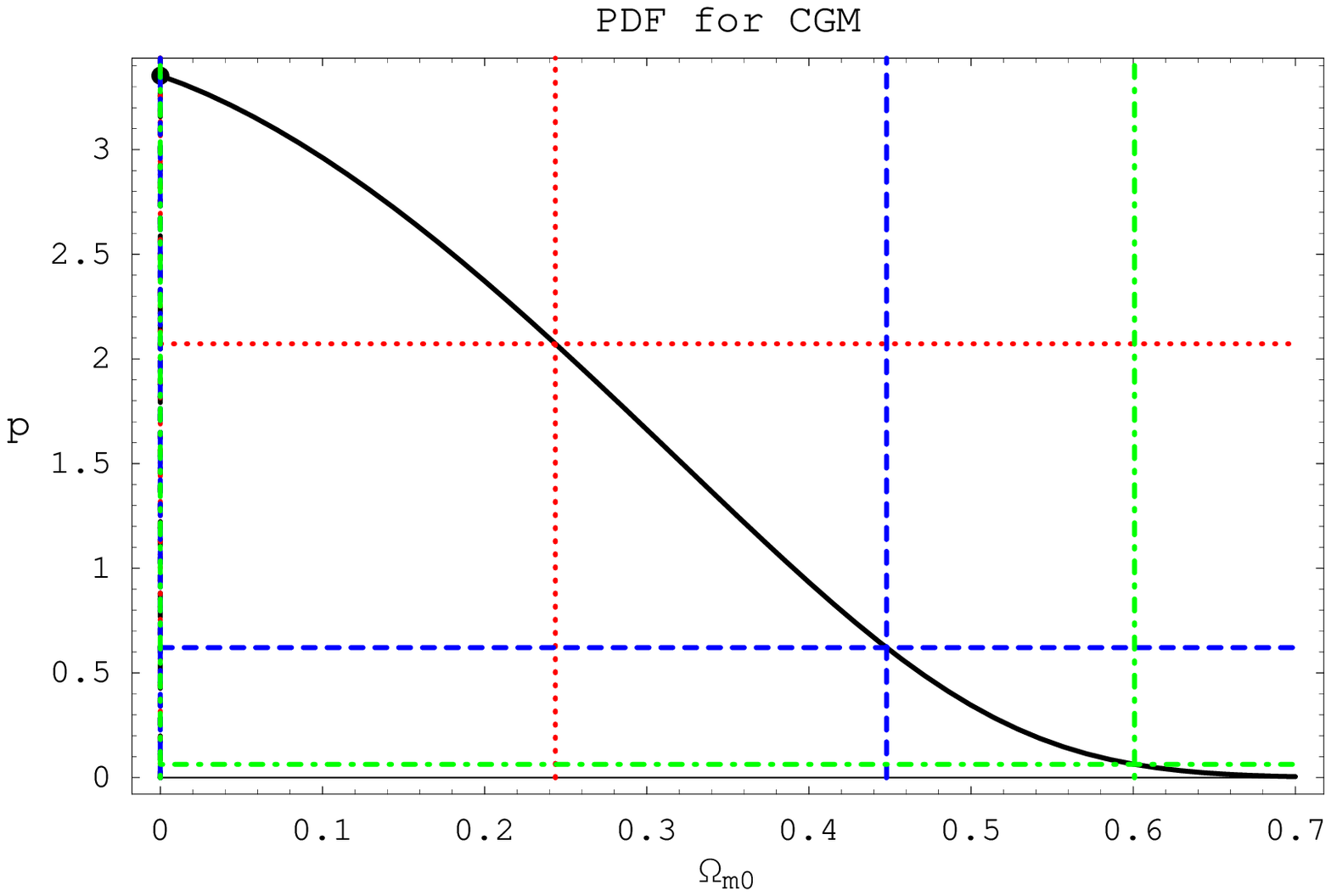}
\end{minipage} \hfill 
\begin{minipage}[t]{0.48\linewidth}
\includegraphics[trim=0.4in 0.2in 0.2in 0in,width=\linewidth]{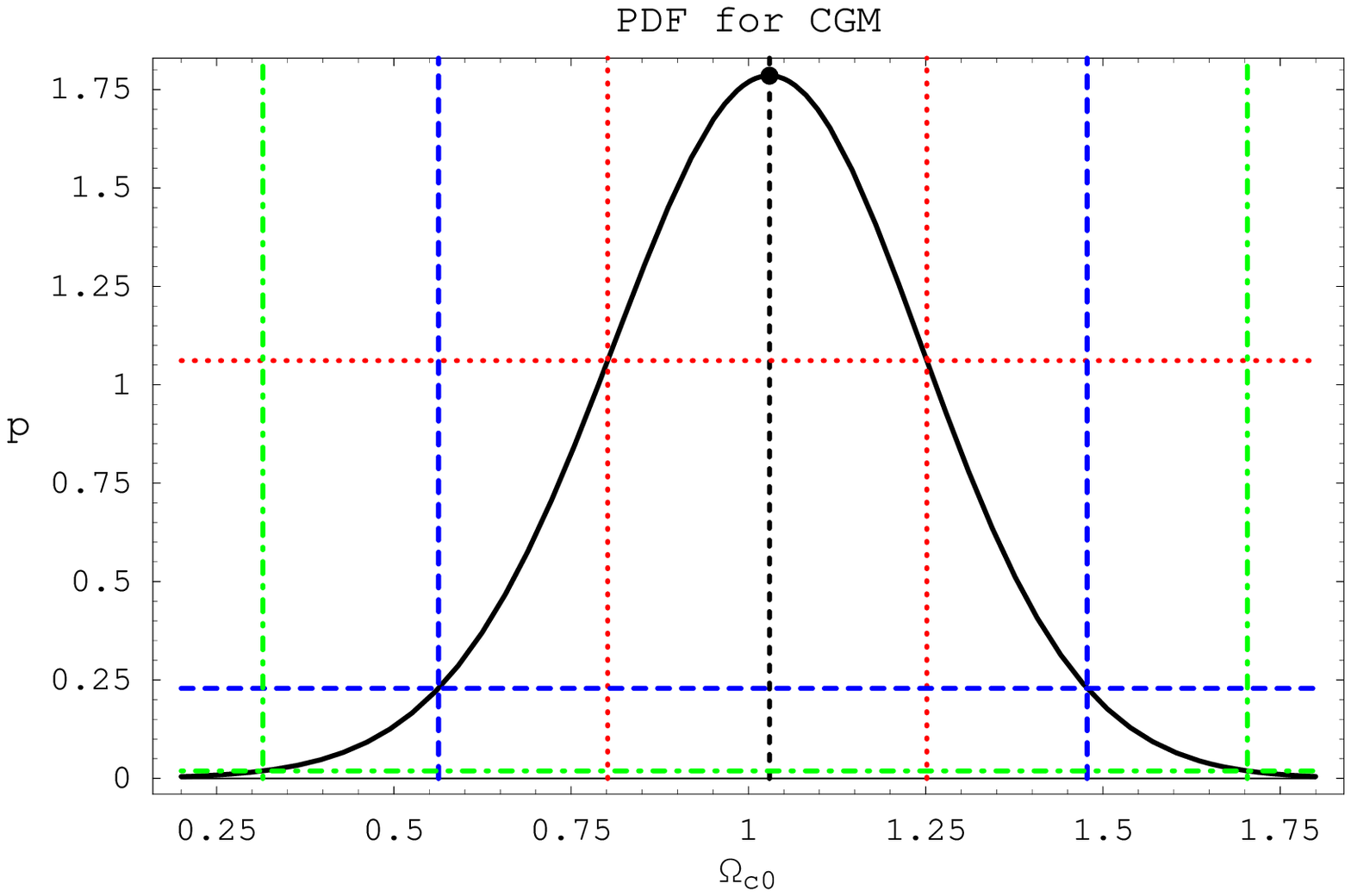}
\end{minipage} \hfill 
\begin{minipage}[t]{0.48\linewidth}
\includegraphics[trim=0.2in 0.2in 0.4in 0in,width=\linewidth]{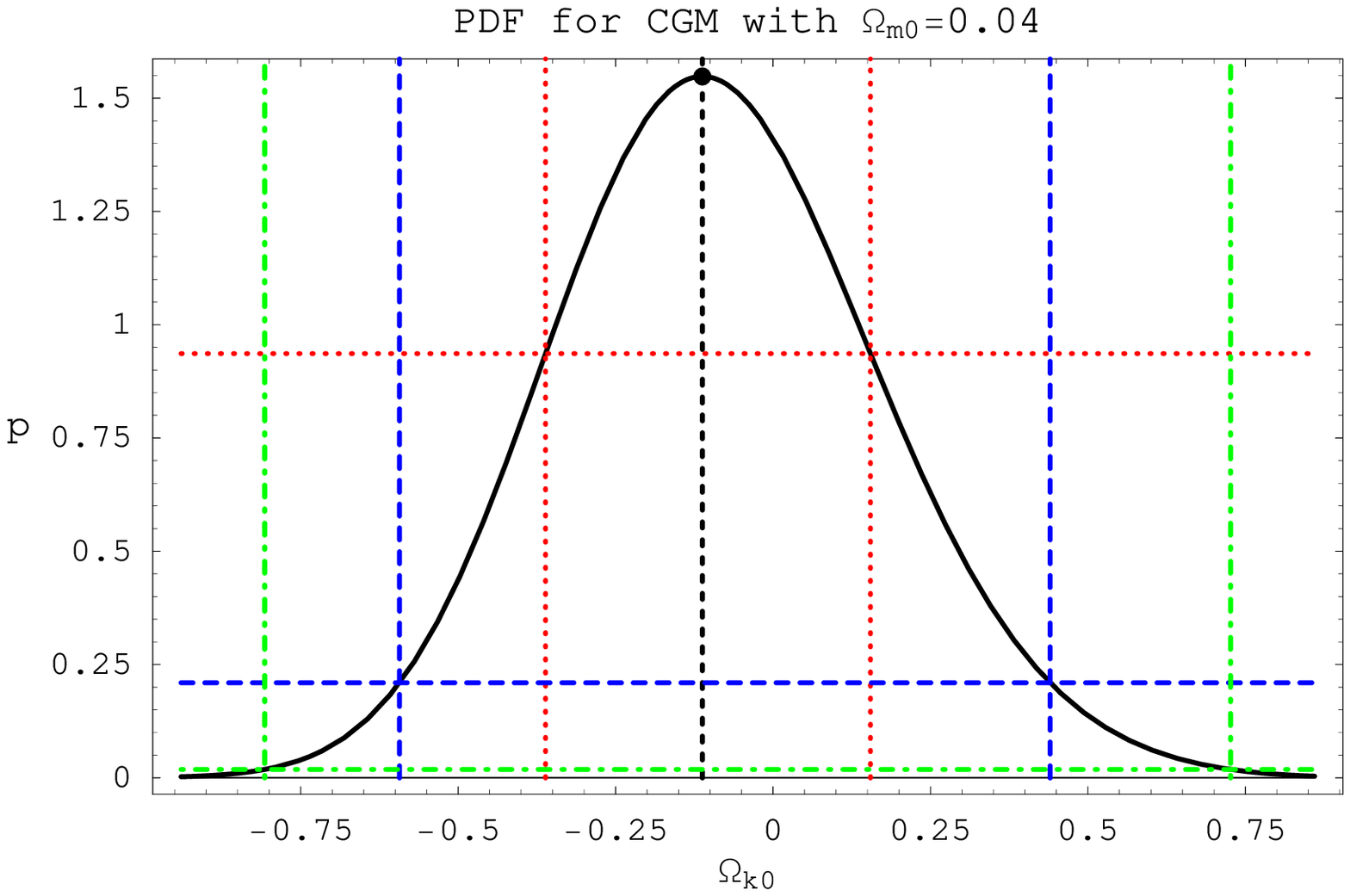}
\end{minipage} \hfill
\caption{{\protect\footnotesize The one-dimensional PDF of $\Omega _{k0}$, $%
\Omega _{m0}$ and $\Omega _{c0}$ for the traditional Chaplygin gas model.
The solid lines are the PDF, the $1\protect\sigma $ ($68.27\%$) regions are
delimited by red dotted lines, the $2\protect\sigma $ ($95.45\%$) credible
regions are given by blue dashed lines and the $3\protect\sigma $ ($99.73\%$%
) regions are delimited by green dashed-dotted lines. The case for $%
\Omega_{m0}=0$ is not shown here because it is similar to the one with $%
\Omega_{m0}=0.04$. The case for $k=0$ is shown in figure $10$ of ref. 
\protect\cite{colistete3}. As $\Omega _{c0}=1-\Omega _{k0}-\Omega _{k0}$,
for $\Omega _{m0}=0$ we have $\Omega _{c0}=1-\Omega _{k0}$, for $\Omega
_{m0}=0.04$ then $\Omega _{c0}=0.96-\Omega _{k0}$ and for $\Omega _{k0}=0$
we also have $\Omega _{c0}=1-\Omega _{m0}$. }}
\label{figsCGOmegas}
\end{figure}

\begin{figure}[ht!]
\begin{minipage}[t]{0.48\linewidth}
\includegraphics[trim=0.4in -0.2in 0.2in 0in,width=\linewidth]{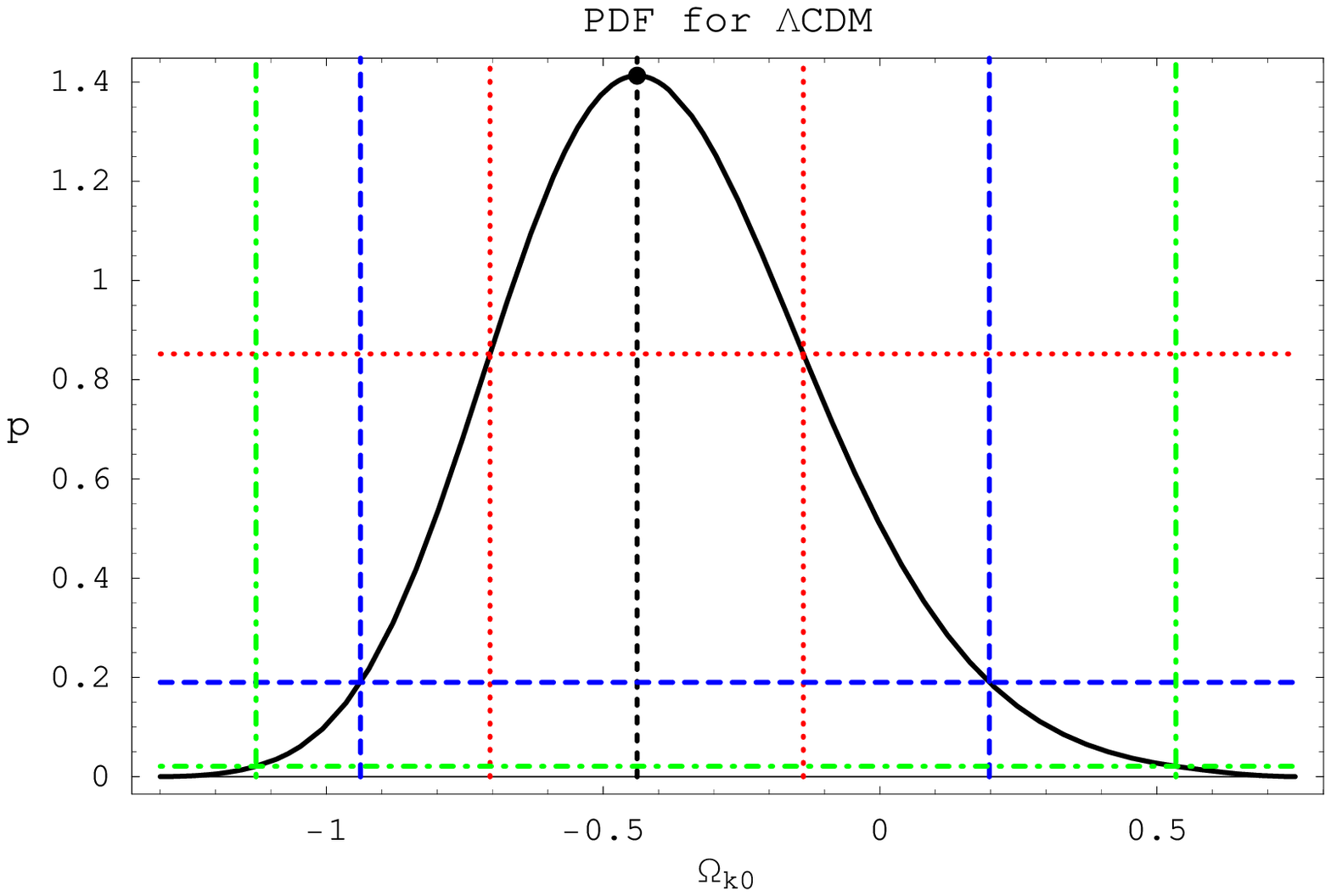}
\end{minipage} \hfill 
\begin{minipage}[t]{0.48\linewidth}
\includegraphics[trim=0.2in -0.2in 0.4in 0in,width=\linewidth]{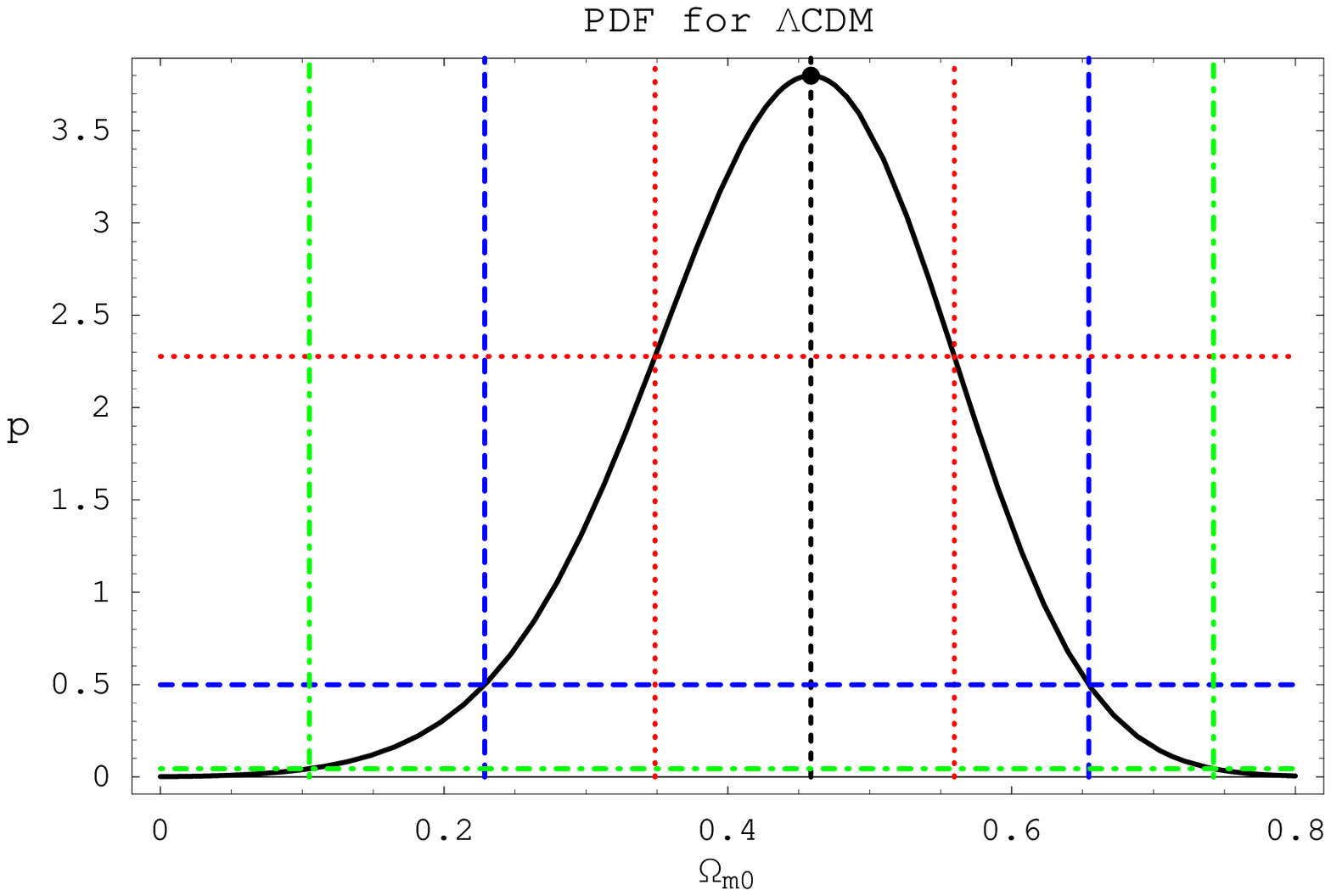}
\end{minipage} \hfill 
\begin{minipage}[t]{0.48\linewidth}
\includegraphics[trim=0.4in 0.2in 0.2in 0in,width=\linewidth]{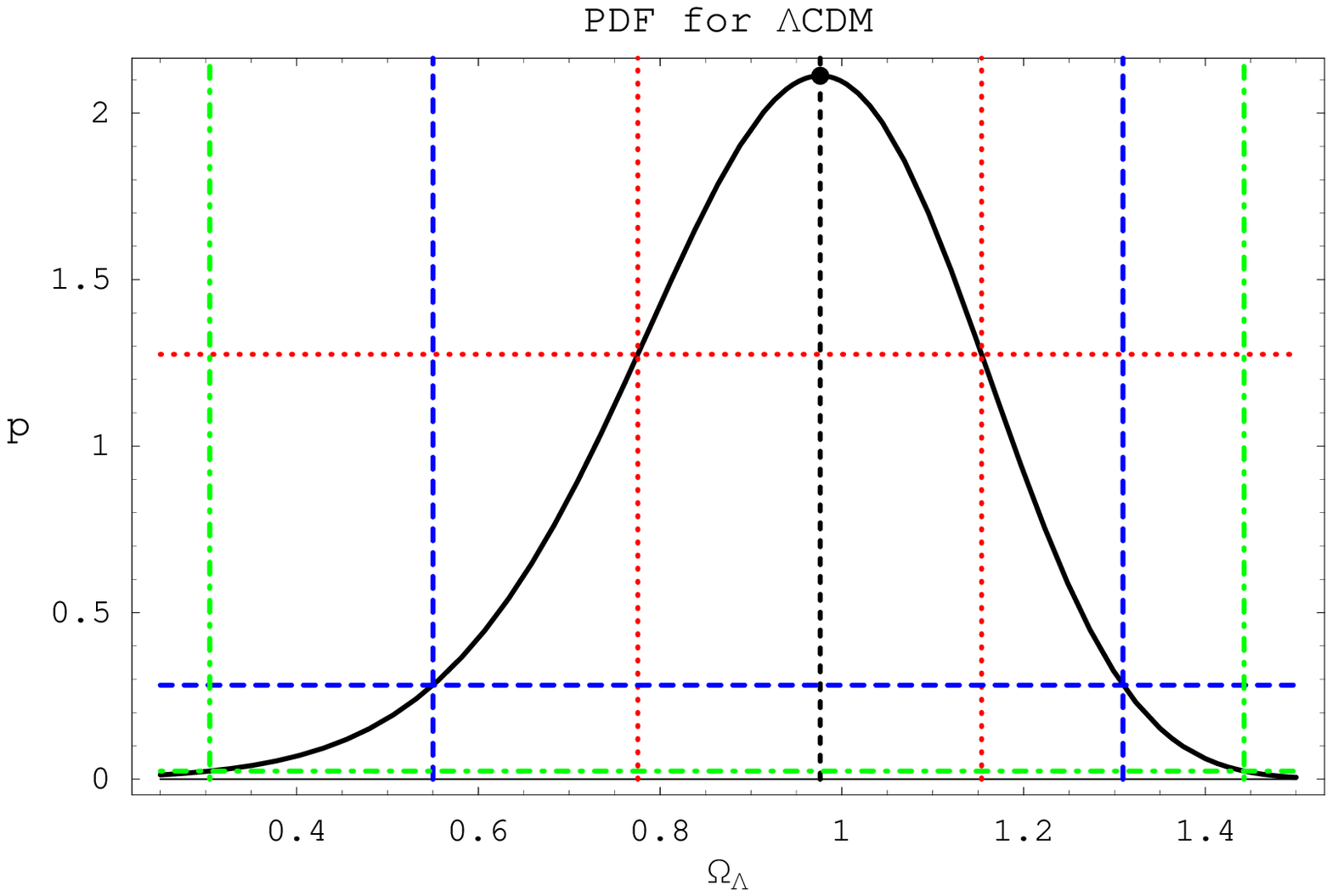}
\end{minipage} \hfill 
\begin{minipage}[t]{0.48\linewidth}
\includegraphics[trim=0.2in 0.2in 0.4in 0in,width=\linewidth]{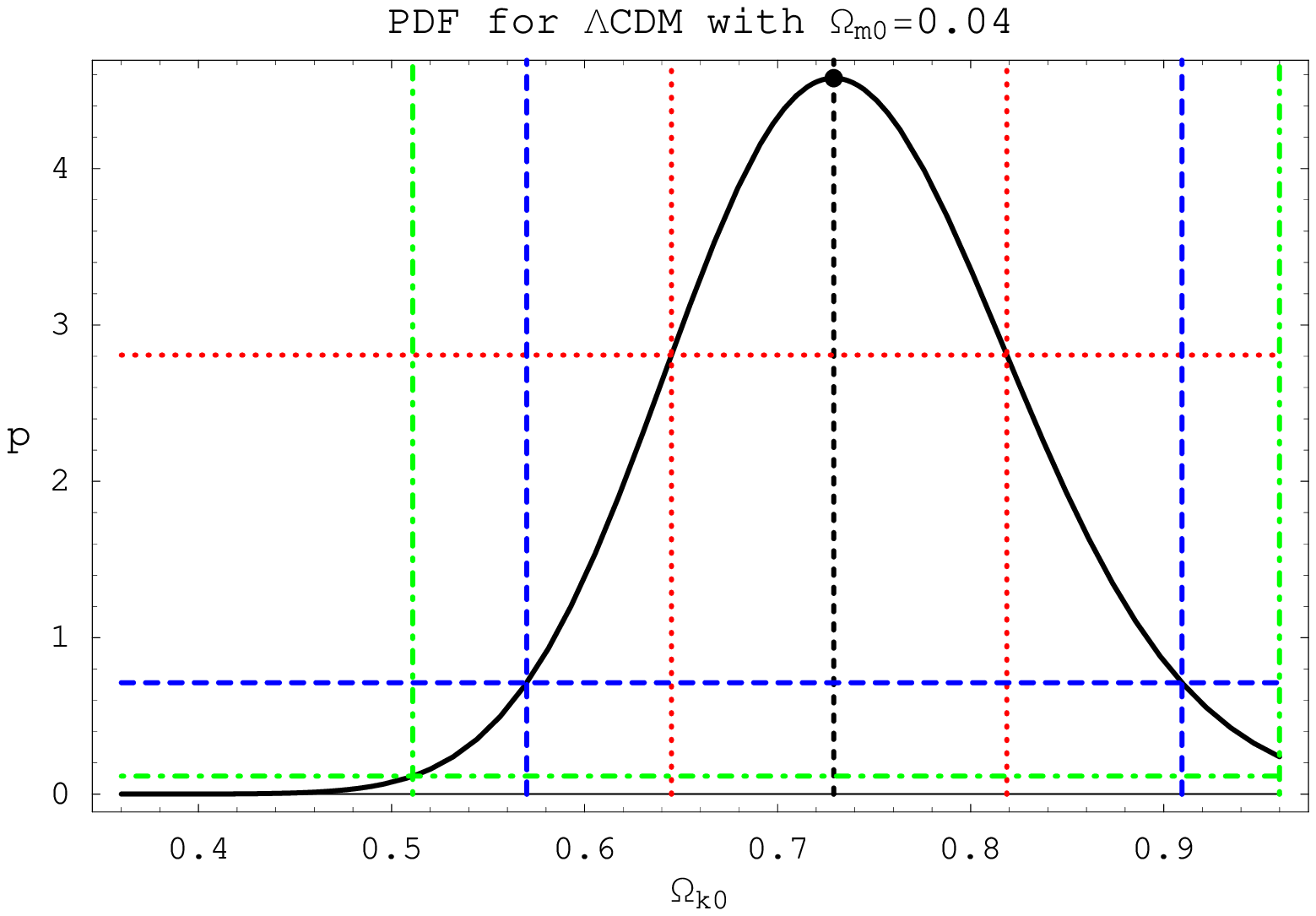}
\end{minipage} \hfill
\caption{{\protect\footnotesize The one-dimensional PDF of $\Omega _{k0}$, $%
\Omega _{m0}$ and $\Omega _{c0}$ for the $\Lambda$CDM model. The solid lines
are the PDF, the $1\protect\sigma $ ($68.27\%$) regions are delimited by red
dotted lines, the $2\protect\sigma $ ($95.45\%$) credible regions are given
by blue dashed lines and the $3\protect\sigma $ ($99.73\%$) regions are
delimited by green dashed-dotted lines. The case for $\Omega_{m0}=0$ is not
shown here because it is similar to the one with $\Omega_{m0}=0.04$. The
case for $k=0$ is shown in figure $11$ of ref. \protect\cite{colistete3}. As 
$\Omega _{c0}=1-\Omega_{k0}-\Omega _{k0}$, for $\Omega _{m0}=0$ we have $%
\Omega _{c0}=1-\Omega_{k0}$, for $\Omega _{m0}=0.04$ then $\Omega
_{c0}=0.96-\Omega _{k0}$ and for $\Omega _{k0}=0$ we also have $\Omega
_{c0}=1-\Omega _{m0}$. }}
\label{figsLCDMOmegas}
\end{figure}

The unified scenario of quartessence, with no pressureless matter, is again
favoured as, for example, the GCGM and CGM (without fixed parameters)
predict $\Omega _{m0}=0.000_{-0.000}^{+0.499}$ and $\Omega
_{m0}=0.000_{-0.000}^{+0.448}$, respectively. The same estimations from ref. 
\cite{colistete2}, $\Omega _{m0}=0.00_{-0.00}^{+0.86}$, and ref. \cite
{colistete1}, $\Omega _{m0}=0.00_{-0.00}^{+0.82}$, show that the increased
number of SNe Ia has substantially decreased the estimated error. Compared
to ref. \cite{colistete3}, the dispersion is also highly decreased, which
once more favours the quartessence scenario. The case of flat Universe has
an even smaller dispersion for the quartessence model. See figures \ref
{figsGCGOmegas} and \ref{figsCGOmegas}.

Like ref. \cite{colistete3}, the GCGM, the CGM and the $\Lambda$CDM
remarkably lead to almost the same predictions concerning the dark energy
component, $\Omega_{c0}$, when all parameters are free: $%
1.012^{+0.667}_{-0.489}$, $1.030^{+0.448}_{-0.467}$ and $%
0.976^{+0.333}_{-0.426}$, respectively. By comparing with ref. \cite
{colistete3}, now $\Omega_{c0}$ (which behaves as $\Omega_{\Lambda}$ for $%
\Lambda$CDM) has lower values and narrower dispersions, like $\Omega_{m0}$,
see figures \ref{figsGCGOmegas}--\ref{figsLCDMOmegas}.

The joint probability for $\Omega_{m0}$ and $\Omega_{c0}$ is now smoother
for the GCGM and the CGM cases (figure \ref{figCGCCGOmegam0c0}), with the $%
1\sigma$, $2\sigma$ and $3\sigma$ contours of figures \ref{figCGCCGOmegam0c0}
and \ref{figLCDMOmegam0c0} showing significantly smaller regions. So the
analysis for $\Lambda$CDM is now quite in agreement with the results of ref. 
\cite{riessa}.

\subsection{Estimation of $\Omega_{k0}$}

In comparison with ref. \cite{colistete3}, a closed Universe is still
clearly favoured, but with slightly smaller probability, i.e., $p(\Omega
_{k0}<0)$. But more important, the dispersion for $\Omega_{k0}$ is now
substantially narrowed, as shown by figures \ref{figsGCGOmegas}--\ref
{figsLCDMOmegas}. The probability to have a flat Universe, $p(\Omega _{k0} =
0)$, is now greater ($45.70\,\%$, $39.84\,\%$ and $15.44\,\%$ for GCGM, CGM
and $\Lambda$CDM), and after setting the pressureless matter it increases,
exception being the $\Lambda$CDM case.

With respect to refs. \cite{colistete1,colistete2} (using the selected 26
SNe Ia data), the dispersion has also significantly decreased, for example : 
$\Omega_{k0}=-0.251_{-0.694}^{+0.605}$ and $%
\Omega_{k0}=-0.228_{-0.508}^{+0.552}$ versus $%
\Omega_{k0}=-0.74_{-1.32}^{+1.42}$ \cite{colistete2} and $%
\Omega_{k0}=-0.84_{-1.23}^{+1.51}$ \cite{colistete1}, for respectively GCGM
and CGM.

\subsection{Estimation of the age of the Universe, $t_0$}

Due to the larger number of SNe Ia used here with respect to refs. \cite
{colistete1,colistete2}, the dispersions have been quite decreased. For
example, $t_0 = 14.42_{-1.77}^{+2.51}\,Gy$ and $t_0 =
13.93_{-0.65}^{+0.97}\,Gy$ estimated here for the GCGM and CGM versus $t_0 =
15.3^{+4.2}_{-3.2}\,Gy$ \cite{colistete2} and $t_0 = 14.2^{+2.8}_{-1.5}\,Gy$ 
\cite{colistete1}, respectively.

The predicted age of the Universe when no parameter is fixed has increased
with respect to ref. \cite{colistete3}, fortunately not anymore dangerously
near the recent estimations age of the globular clusters \cite{krauss}, $t_0
= 12.6^{+3.4}_{-2.4}\,Gy$.

\subsection{Estimation of the deceleration parameter $q_0$}

The values for the deceleration parameter $q_0$ are increased (less
negative) with respect to ref. \cite{colistete3}. The estimated errors are
significantly smaller than the ones of refs. \cite{colistete1,colistete2},
for example : $q_{0}=-0.730_{-0.328}^{+0.352}$ and $%
q_{0}=-0.769_{-0.309}^{+0.358}$ versus $q_{0}=0.80_{-0.62}^{+0.86}$ \cite
{colistete2} and $q_{0}=0.98_{-0.62}^{+1.02}$ \cite{colistete1}, for
respectively GCGM and CGM.

In all cases, $p(q_{0}<0)$, the probability to have an accelerating Universe
today, is equal to or very near $100\%$.

\subsection{Estimation of the scale factor $a_{i}$ the Universe begins to
accelerate from}

Another useful quantity is the scale factor at the moment the Universe
begins to accelerate, $a_{i}$, keeping in mind that the scale factor is
normalized with its present value $a_{0}$ equal to unity. With respect to
ref. \cite{colistete3}, $a_{i}$ decreases when no parameter is fixed. As
expected, the larger number of supernovae in comparison with ref. \cite
{colistete2} yields smaller credible intervals for $a_{i}$, for example $%
a_{i}=0.626_{-0.123}^{+0.184}\,a_{0}$ estimated here for GCGM versus $%
a_{i}=0.67_{-0.37}^{+0.25}\,a_{0}$ of ref. \cite{colistete2}.

The probability the Universe begins to accelerate before today, $p(a_{i}<1)$
in tables \ref{tableParEstGCG}-- \ref{tableParEstLCDMp}, is essentially $%
100\%$, being approximately the same value of the probability to have an
accelerating Universe today, i.e., $p(q_{0}<0)$. Theoretically they should
be the same, so the fact that these independent probability calculations
agree almost exactly shows the accuracy and reliability of the Bayesian
probability analyses of this work.

\section{Conclusions}

\label{sectionConclusions}

The present work has performed the most extensive analysis of \ the GCGM and
CGM in what concerns the comparison of theoretical predictions with the type
Ia supernovae data, using the 157 data of the ``gold sample''. By using the
high productivity of \textbf{BETOCS} (\textbf{B}ay\textbf{E}sian \textbf{T}%
ools for \textbf{O}bservational \textbf{C}osmology using \textbf{S}Ne Ia 
\cite{betocs}), it was possible to make all the best-fittings, parameter
estimations and figures shown here.

All positive features of GCGM and CGM were enhanced with respect to ref. 
\cite{colistete3} :

\begin{itemize}
\item  the probability to have $\alpha >0$ (with more physical meaning) is
increased, as well as both $p(\alpha =1)$ and $p(\alpha >0)$;

\item  value of $\bar{A}$ now has slightly smaller dispersion;

\item  the dispersion of $\Omega _{m0}$ and $\Omega _{c0}$ are also highly
decreased, favouring even more the quartessence \cite{maklerthesis,
makler2003} scenario;

\item  the dispersion for $\Omega _{k0}$ is now substantially narrowed, and
the probability to have a flat Universe is now greater;

\item  predicted age of the Universe (when no parameter is fixed) has
increased, fortunately not anymore dangerously near the recent estimations
age of the globular clusters.
\end{itemize}

One important result concerns the Hubble parameter, $H_{0}$. It is not
acceptable to fix its value because arbitrary parameter estimations and
usually bad best-fittings are obtained. As the HST (Hubble Space Telescope)
prior \cite{freedman} for $H_{0}$ implies minor effects on all results, we
can choose a flat or HST prior.

The CGM (traditional Chaplygin gas model), where $\alpha =1$, remains
competitive and preferred in many cases : when the five parameters are
considered, the probability is $40.81\%$, but increases as much as to $%
95.94\%$ for the quartessence scenario.

For the parameter $\bar{A}$, both GCGM and CGM cases of fixed curvature and
matter densities shows a value near but small than $1$ as the best value of $%
\bar{A}$, such that the $\Lambda $CDM case ($\bar{A}=1$) is almost ruled out.

The results indicate that, for the GCGM, CGM and $\Lambda $CDM, a closed
Universe is favoured. The GCGM and CGM favour the unified scenario
(quartessence), where the pressureless matter density is essentially zero.

There are many current and future applications \cite{colistete3, colistete1,
colistete2, cfgv} and developments of BETOCS : using different SNe Ia data
sets (SNLS \cite{snls}, etc), different priors, different cosmological
models \cite{nesseris2004}, etc. We plan to release variant versions for
other observational cosmological data \cite{jassal, nesseris2006, cunha,
amendola}, like \textbf{BETOCX} (\textbf{B}ay\textbf{E}sian \textbf{T}ools
for \textbf{O}bservational \textbf{C}osmology using \textbf{X}-ray gas mass
fraction of galaxy clusters), so it will be possible to cross the
estimations from different observational data.

\vspace{0.5cm} \noindent \textbf{Acknowledgments} \newline
\newline
\noindent The authors are also grateful by the received financial support
during this work, from FACITEC\-/\-PMV (R. C. Jr.) and CNPq (R. G.).


\begin{thebibliography}{99}
\bibitem{riess}  A. Riess et al, Astron. J. \textbf{116}, 1009 (1998);

\bibitem{mutter}  S. Perlmutter et al, Astrophys. J. \textbf{517}, 565
(1999);

\bibitem{spergel}  D. N. Spergel et al, Astrophys. J. Suppl. \textbf{148},
175 (2003);

\bibitem{turner}  E. L. Turner, J. P. Ostriker, and J. R. Gott III,
Astrophys. J. \textbf{284}, 1 (1984).

\bibitem{fukugita}  M. Fukugita, T. Futamase, and M. Kasai, Mont. Not. R.
Astron. Soc. \textbf{246}, 24 (1990).

\bibitem{allen04}  S.W. Allen \textit{et al.}, Monthly Notices of the Royal
Astron. Society \textbf{353}, 457 (2004);

\bibitem{bagla}  Bagla J. S., Padmanabhan T., Narlikar J. V., 1996, Comments
on Astrophysics 18, 275; astro-ph/9511102;

\bibitem{weinberg}  S. Weinberg, Rev. Mod. Phys. \textbf{61}, 1 (1989);

\bibitem{carroll}  S. M. Carroll, Living Rev. Rel. \textbf{4}, 1 (2001);

\bibitem{stein1}  R. R. Caldwell, R. Dave and P. J. Steinhardt, Phys. Rev.
Lett. \textbf{80}, 1582 (1998);

\bibitem{stein2}  I. Zlatev, L. Wang and P. J. Steinhardt, Phys. Rev. Lett. 
\textbf{82}, 896 (1999);

\bibitem{kamenshchik}  A. Kamenshchik, U. Moschella, and V. Pasquier, Phys.
Lett. B \textbf{511}, 265 (2001);

\bibitem{maklerthesis}  M. Makler, \textit{Gravitational Dynamics of
Structure Formation in the Universe}, PhD Thesis, Brazilian Center for
Research in Physics (2001);

\bibitem{bilic2002}  N. Bili\'{c}, G.B. Tupper, and R.D. Viollier, Phys.
Lett. B \textbf{535}, 17 (2002);

\bibitem{bento2002}  M.C. Bento, O. Bertolami, and A.A. Sen, Phys. Rev. D 
\textbf{66}, 043507 (2002);

\bibitem{jack}  R. Jackiw, \textit{A particle field theorist's lectures on
supersymmetric, non-abelian fluid mechanics and d-branes}, physics/0010042;

\bibitem{fabris}  J. C. Fabris, S. V. B. Gon\c{c}alves and P. E. de Souza,
Gen. Rel. Grav. \textbf{34}, 53 (2002);

\bibitem{sand}  H. Sandvik, M. Tegmark, M. Zaldarriaga e I. Waga, Phys. Rev. 
\textbf{D69}, 123524 (2004);

\bibitem{avelino}  L. M. G. Be\c{c}a, P. P. Avelino, J. P. M. de Carvalho
and C. J. A. P. Martins, Phys. Rev. \textbf{D67}, 101301 (2003);

\bibitem{colistete3}  R. Colistete Jr., J. C. Fabris, Class. Quantum Grav. 
\textbf{22}, 2813 (2005);

\bibitem{colistete1}  R. Colistete Jr., J. C. Fabris, S. V. B. Gon\c{c}alves
and P. E. de Souza, Int. J. Mod. Phys. \textbf{D13}, 669 (2004);

\bibitem{colistete2}  R. Colistete Jr., J. C. Fabris and S. V. B. Gon\c{c}%
alves, Int. J. Mod. Phys. \textbf{D14}, 775 (2005);

\bibitem{betocs}  R. Colistete Jr., \textit{\textbf{B}ay\textbf{E}sian 
\textbf{T}ools for \textbf{O}bservational \textbf{C}osmology using \textbf{S}%
Ne (\textbf{BETOCS})}, available on the Internet site
http://www.RobertoColistete.net/BETOCS, (2006);

\bibitem{mathematica}  S. Wolfram, \textit{The Mathematica Book} (Cambridge
University Press), \ 1999.

\bibitem{freedman}  W. Freedman, Astrophys. J. \textbf{553}, 47 (2001).

\bibitem{riessa}  A. G. Riess et al, Astrophys. J. \textbf{607}, 665 (2004);

\bibitem{makler2003}  M. Makler, S. Q. Oliveira, and I. Waga, Phys. Lett. 
\textbf{B}, \textbf{555}, 1 (2003).

\bibitem{Loredo}  T. J. Loredo, in `Maximum Entropy and Bayesian Methods',
ed. P.F. Fougere, Kluwer Academic Publishers, Dordrecht (1990) ;

\bibitem{Loredo2}  T. J. Loredo, in `Statistical Challenges in Modern
Astronomy', ed. E.D. Feigelson and G.J. Babu, Springer-Verlag, New York
(1992);

\bibitem{Gregory}  P. C. Gregory and T.J. Loredo, Astrophys. J. \textbf{398}%
, 146 (1992);

\bibitem{Abroe}  M. E. Abroe et al., Mon. Not. R. A. Soc. \textbf{334}, 11
(2002);

\bibitem{krauss}  L.M. Krauss, \textit{The state of the Universe:
cosmological parameters 2002}, astro-ph/0301012;

\bibitem{cfgv}  F. Casarejos, J. C. Fabris, S. V. B. Gon\c{c}alves and J. F.
Villas da Rocha, \textit{Constraining double component dark energy model
using type Ia supernovae data}, astro-ph/0606171;

\bibitem{snls}  P. Astier et al. (SNLS colaboration), Astron. Astrophys. 
\textbf{447}, 31 (2006);

\bibitem{nesseris2004}  S. Nesseris and L. Perivolaropoulos, Phys. Rev. D 
\textbf{70}, 043531, (2004);

\bibitem{jassal}  H. K. Jassal, J. S. Bagla and T. Padmanabhan, \textit{The
vanishing phantom menace}, astro-ph/0601389;

\bibitem{nesseris2006}  S. Nesseris and L. Perivolaropoulos, \textit{%
Crossing the Phantom Divide: Theoretical Implications and Observational
Status}, astro-ph/0610092;

\bibitem{cunha}  J. V. Cunha, J. S. Alcaniz and J. A. S. Lima, Phys. Rev. 
\textbf{D69}, 083501 (2004);

\bibitem{amendola}  L. Amendola, M. Makler, R. R. R. Reis and I. Waga, Phys.
Rev. D \textbf{74}, 063524 (2006).
\end{thebibliography}
\end{document}